
\input phyzzx.tex
\catcode`\@=11
\def\Smallpoint{\rel@x
\def\Tinpoint{\tenpoint \rel@x
\ifsingl@\subspaces@t2:5;\else\subspaces@t3:5;\fi
\ifdoubl@ \multiply\baselineskip by 5
\divide\baselineskip by 4\fi}
\parindent=16.67pt
\itemsize=25pt
\thinmuskip=2.5mu
\medmuskip=3.33mu plus 1.67mu minus 3.33mu
\thickmuskip=4.17mu plus 4.17mu
\def\thinspace{\kern .13889em}
\def\negthinspace{\kern-.13889em}
\def\enspace{\kern.416667em}
\def\enskip{\hskip.416667em\rel@x}
\def\quad{\hskip.83333em\rel@x}
\def\qquad{\hskip1.66667em\rel@x}
\def\crr{\cropen{8.3333pt}}
\def\papersize{\vsize=39.0pc\hsize=28.6pc
\hoffset=-.55truein\voffset=.7pc
\skip\footins\bigskipamount}
\Tenpoint\papersize}
\catcode`\@=12
\newbox\leftpage
\newdimen\fullhsize
\newdimen\hstitle
\newdimen\hsbody
\def\LITTLE{\Smallpoint\relax
\let\lr=L
\hstitle=8truein\hsbody=4.75truein\fullhsize=10truein\hsize=\hsbody
\output={\ifnum\pageno=0
  \shipout\vbox{\hbox to \fullhsize{\hfill\pagebody\hfill}}\advancepageno
  \else
  \almostshipout{\leftline{\vbox{\pagebody\makefootline}}}\advancepageno
  \fi}
\def\almostshipout##1{\if L\lr
      \global\setbox\leftpage=##1 \global\let\lr=R
  \else
    \shipout\vbox{
      \hbox to\fullhsize{\box\leftpage\hfil##1}}  \global\let\lr=L\fi}
}
\LITTLE

\pagenumbers

 ~
\vskip -0.1 truein
\date{McGill/91-29;~~CLNS 91/1100 \cr
November 1991 \cr}
\titlepage
\title{\bf MODEL-BUILDING FOR FRACTIONAL SUPERSTRINGS}
\author{Keith R. Dienes\foot{E-mail address:  dien@hep.physics.mcgill.ca}}
\address{Dept.~of Physics, McGill University\break
E.~Rutherford Building, 3600 University St.\break
Montr\'eal, P.Q.~H3A-2T8 ~~Canada}
\andauthor{S.-H. Henry Tye}
\address{Newman Laboratory of Nuclear Studies \break
Cornell University \break
Ithaca, NY  14853-5001 ~~USA}
\bigskip
\abstract
We investigate heterotic string-type model-building for
the re\-cently-pro\-posed frac\-tional super\-string theories.
We concentrate on the cases with critical spacetime dimensions
four and six, and find that a correspondence can be drawn
between the new fractional superstring models and a special subset
of the traditional heterotic string models.  This allows us to generate
the partition functions of the new models, and demonstrate that their
number is indeed relatively limited.  It also appears that these strings
have uniquely natural compactifications to lower dimensions.
In particular, the $D_c=6$ fractional superstring has a natural
interpretation in four-dimensional spacetime.
\endpage

\def\calZ{{\cal Z}}
\def\calF{{\cal F}}
\def\calK{{\cal K}}
\def\bff{{\bf f}}
\def\bZ{{\bf Z}}
\def\calS{{\cal S}}
\def\half{{\textstyle {1\over2}}}
\def\quarter{{\textstyle {1\over4}}}
\def\Nmax{{N_{\rm max}}}
\def\ie{{\it i.e.}}
\def\eg{{\it e.g.}}
\def\equalstar{{ \buildrel \ast \over = }}
\def\W#1{{\bf W_{#1} }}
\def\tautwo{{ \tau_2}}
\def\tauone{{ \tau_1}}
\def\qbar{{  \overline{q} }}
\def\thetafour{{ \vartheta_4 }}
\def\thetatwo{{ \vartheta_2 }}
\def\thetathree{{ \vartheta_3 }}

\def\varthetabar{{ \overline{\vartheta} }}

\def\deltaminusone {{   \Delta^{-1}  }}
\def\deltabarminushalf{{  {\overline{\Delta}}^{-1/2}  }}
\def\deltaminushalf{{  \Delta^{-1/2}  }}

\def\deltahalf{{  \Delta^{1/2}  }}

\def\ket#1{{ |{#1}\rangle }}
\def\bbar{{\overline{\beta}}}
\def\cbar{{\overline{\gamma}}}
\def\dbar{{\overline{\delta}}}
\def\Jbar{{\overline{J}}}

\hyphenation{pa-ra-fer-mion pa-ra-fer-mion-ic pa-ra-fer-mions }
\hyphenation{su-per-string frac-tion-ally su-per-re-pa-ra-met-ri-za-tion}
\hyphenation{su-per-sym-met-ric frac-tion-ally-su-per}
\hyphenation{space-time-super-sym-met-ric fer-mi-on}

\doublespace
\chapter{INTRODUCTION}

It is generally accepted that
string theory is the only known theory offering a realistic
hope of unifying all forces and matter in nature.  In particular,
it provides an attractive solution to the problem of reconciling
quantum mechanics and general relativity.  To date,
the only known consistent string theories are the superstring
theory\REF\GSW{See, \eg, M.B.~Green, J.H.~Schwarz and
E.~Witten, {\it Superstring Theory}, Vols.~1 and 2 (Cambridge University
Press, 1987);
M.~Kaku, {\it Introduction to Superstrings} (Springer-Verlag, 1988).}\refend\
and the closely-related heterotic string theory,\REF\heteroticrefs{
D.J.~Gross, J.A.~Harvey, E.~Martinec, and
R.~Rohm, {\it Nucl.~Phys.} {\bf B256}, 253 (1985).}\refend\
yet one of the significant problems
involved in superstring model-building
in four spacetime dimensions is the existence of
a multitude of classical solutions which these string
theories allow.  This
lack of uniqueness arises because these theories
have critical dimension $D=10$;
in order to have a sensible interpretation
in four spacetime dimensions,
one must therefore choose an arbitrary compactification
scheme for these six extra dimensions or
represent these extra degrees of freedom in terms of arbitrarily
chosen additional worldsheet fields.
There are indeed many ways in which this can be done,
resulting in many millions of possible classical vacua.
It is presumed that some dynamical mechanism or symmetry
argument might be used to select the vacuum corresponding
to the present-day physical world, but at present such
approaches have not been successful.

Recently, however, a new approach to superstring theory
and superstring model-building has been
proposed.\REF\letter{P.C.~Argyres and
S.-H.H.~Tye, {\it Fractional Superstrings with Space-Time Critical
Dimensions Four and Six}, Cornell preprint CLNS 91/1068 (1991),
to appear in {\it Phys.~Rev.~Lett}.}\refend\
Rather than work within the superstring/heterotic framework,
the fundamental idea is to construct a new type of string theory
called the {\it fractional superstring}
in which four can appear as the critical spacetime dimension directly.
Such a theory would therefore not only serve as a more natural
starting-point for describing our physical world, but hopefully also
lead to a smaller and thereby more compelling
set of self-consistent classical vacua.
The basic idea behind the fractional superstring is to replace the
traditional right-moving worldsheet
supersymmetry found in heterotic strings with
a right-moving worldsheet {\it fractional}\/ supersymmetry
parametrized by an integer $K_R\geq 2$.
Such a fractional supersymmetry relates worldsheet bosons not to fermions
but rather to worldsheet {\it parafermions}, and
one finds that the corresponding critical dimension of such a string theory is
$$  D ~=~ \cases{
  2~+~ 16/K_R & for $K_R\geq 2~~$ (fractional supersymmetry),\cr
  26 & for $K_R=1~~$ (no supersymmetry). \cr}~
   \eqn\dimensionformula
$$
The case $K_R=1$ ($D=26$) corresponds to the traditional bosonic
string, and $K_R=2$ ($D=10$) corresponds to the traditional super- or
heterotic string.  The new theories are those for which $K_R>2$, and
we see that by choosing the cases
$K_R=8$ ($D=4$) or $K_R=4$ ($D=6$),
we can obtain significantly lower critical dimensions.
For these cases we can therefore expect a smaller set of classical vacua
and hopefully a more natural description of the physical world.

These fractional superstring theories are indeed natural extensions
of the $K_R=2$ superstring theories, and it is straightforward
to relate the two in terms of their underlying worldsheet physics.
For the traditional superstring theory,
it is well-known that the underlying worldsheet structure
is closely related to the $SU(2)_2$ Wess-Zumino-Witten (WZW)
theory:\REF\Knizh{
V.G.~Knizhnik and A.B.~Zamolodchikov, {\it Nucl.~Phys.} {\bf B247}, 83
(1984).}\refend\
the worldsheet superpartner of the spacetime coordinate $X^\mu$
is a Majorana fermion $\psi^\mu$, and this fermionic theory can be
simply described by the WZW coset
$SU(2)_2/U(1)$.  Setting the remaining $U(1)$ boson to an
appropriately chosen radius reproduces the full $SU(2)_2$ WZW theory,
yet if we relax the
radius of this $U(1)$ boson to infinity, we can interpret this boson
as the spacetime coordinate $X^\mu$.
As expected, this decompactification procedure destroys the $SU(2)$
symmetry of the WZW model, but its superconformal symmetry survives
and exists in the worldsheet theory.
The {\it fractional}\/ superstring theory has a similar worldsheet
structure, and is related in precisely the same way to the more general
$SU(2)_K$ WZW theories for $K\geq 2$.
The general coset theories $SU(2)_K/U(1)$
are simply the $\bZ_K$ parafermion theories,\REFS\Zam{A.B.~Zamolodchikov and
V.A.~Fateev, {\it Sov.~Phys. J.E.T.P.} {\bf 62}, 215
(1985).}\REFSCON\Gep{D.~Gepner and Z.~Qiu, {\it Nucl.~Phys.} {\bf B285},
423  (1987).}\refsend\
and we once again obtain our spacetime coordinate field $X^\mu$
by decompactifying the remaining WZW $U(1)$ boson.
Replacing our usual supercurrent for $K>2$ is a new
current\REF\Bag{
D.~Kastor, E.~Martinec, and Z.~Qiu, {\it Phys.~Lett.} {\bf
200B}, 434 (1988);  J.~Bagger, D.~Nemeschansky, and
S.~Yankielowicz, {\it Phys.~Rev.~Lett.} {\bf 60}, 389 (1988);
F.~Ravanini, {\it Mod.~Phys.~Lett.} {\bf A3}, 397 (1988).
See also S.-W.~Chung, E.~Lyman, and S.-H.H.~Tye, {\it Fractional
Supersymmetry and Minimal Coset Models in Conformal Field Theory},
Cornell preprint CLNS 91/1057  (1991).  }\refend\
whose conformal dimension (or equivalently, spin) is $(K+4)/(K+2)$;
these new currents therefore have fractional spin, and
(as we shall see) transform $X^\mu$ to the fractional-spin
field $\epsilon^\mu$, the energy operator in the parafermion theory.
It is therefore natural to refer to this remaining worldsheet symmetry
as a fractional superconformal symmetry, and to the strings based
on these worldsheet symmetries as fractional superstrings.
Because the corresponding fractional superconformal algebra
is non-local on the worldsheet,\REF\Jim{P.C.~Argyres, J.~Grochocinski,
and S.-H.H.~Tye, {\it Nucl.~Phys.~}{\bf B367}, 217 (1991).}\refend\
the analysis for such a string
theory is substantially more involved than for the simpler, local,
superconformal case;  however, as we shall see,
concrete progress can indeed be made.

In this paper we shall assume the underlying consistency of the fractional
superstring and concern ourselves primarily with examining the space
of models that such fractional string
theories allow;  other important issues, such as understanding
the underlying Fock-space structure of these theories,
constructing scattering amplitudes,
examining the ghost system and developing a no-ghost theorem,
are actively being pursued.
Our approach, therefore, is to examine this space of models by
studying the allowed partition functions that such models might have;
in this way we are able to obtain a number of interesting results.

First, we demonstrate that the fractional superstring
partition functions are straightforward generalizations
of the traditional $(K_L,K_R)=(2,2)$ superstring and
$(1,2)$ heterotic string partition
functions, and we explicitly develop a general procedure for
constructing modular-invariant partition functions for our new models
which are consistent with $N=1$ spacetime supersymmetry.

Second, and perhaps more importantly, by focusing much of
our attention on the $(K_L,K_R)=
(1,8)$ and $(1,4)$ heterotic-type fractional superstring models,
we find that we are able to make direct correspondences between
these models and the traditional $(1,2)$ models;  these correspondences
are possible because both theories are built from identical bosonic
left-moving sectors.
These correspondences afford us a means of generating what we believe
to be valid $(1,K_R)$-type fractional superstring models, and we present
a number of concrete examples with critical spacetime dimensions four and six.

Third, our correspondences suggest
that only traditional $(1,2)$ models with a {\it maximal}\/ number
of spacetime supersymmetries can be related to fractional $(1,K_R)$ models
(which were themselves constructed with $N=1$ spacetime supersymmetry).
This result therefore severely constrains the space of fractional
superstring models in $D=4$ and $D=6$, confirming our expectation
that the number of allowed models is indeed relatively small.

Finally, we discuss an intriguing feature
wholly new to these fractional superstring theories:
quantum mechanics, locality, and
Lorentz-invariance together seem to intrinsically select
certain ``natural'' spacetime dimensions
in which these fractional superstring theories must be interpreted.
These ``natural'' dimensions are even smaller than
their critical dimensions, yet we find that
our theories themselves seem to
induce such compactifications in order to achieve Lorentz invariance.
These compactifications, unlike those of the traditional
string theories, are not at all arbitrary, and
we find in particular that the $K=4$ fractional superstring
has a ``natural'' interpretation in four-dimensional spacetime.
It turns out that this compactification may also simultaneously
afford us with a means of building models containing chiral fermions
in fundamental representations of relevant gauge groups.

All of our results therefore not only lend credence to the
frac\-tion\-al-su\-per\-string idea, but may also, we hope, serve as the
first step towards a rigorous model-building program.
In particular, the correspondences we develop here arise from
very general principles, and thus should extend naturally to
a variety of $(K_L,K_R)$ fractional superstring theories
in either their critical or ``natural'' dimensions.

Our goals in this paper are two-fold:
not only do we present the new results discussed
above, but we also aim to provide
a more detailed exposition of the original fractional
superstring idea than was given in Ref.~[\letter].
Accordingly, this paper is somewhat lengthy;  its organization is as follows.
In Sect.~II we
provide a self-contained introduction to the fractional superstring
idea, demonstrating how this approach forms a natural generalization
of the traditional superstring approach.  We then
in Sect.~III
proceed to survey the algebraic forms we expect partition functions
to have for a general $(K_L,K_R)$ string theory:  for $K=1,2$
we present known models which will play a role in later sections,
and for other values of $K$ we introduce the parafermionic
string functions\REF\kacpeterson{
V.G.~Ka\v{c}, {\it Adv.~Math.} {\bf 35}, 264 (1980);
V.G.~Ka\v{c} and D.~Peterson, {\it Bull.~AMS} {\bf 3}, 1057
(1980); {\it ibid.,} {\it Adv.~Math.} {\bf 53}, 125
(1984).}\refend\
and discuss how they enter into the new total partition
functions of $(K_L,K_R)$ models.  We also demonstrate that spacetime
supersymmetry can be incorporated for these models by choosing
these string functions in certain linear combinations, and present
a number of important new string-function identities.
In Sect.~IV we then turn our attention to the heterotic $(1,K_R)$
theories, ultimately deriving various ``dictionaries'' relating these models
to the traditional $(1,2)$ models.  We illustrate the use
of these dictionaries by obtaining a number of new fractional-superstring
models in $D=4$ and $D=6$, and in Sect.~V we
discuss precisely which traditional models may be ``translated''
with these dictionaries.  In this way we observe an expected truncation
in the size of the space of fractional superstring models
relative to that corresponding to traditional superstring
models in $D<10$, reflecting the fact that these
new models are indeed {\it in} their critical dimensions.
We close in Sect.~VI with our discussion of various
further issues in fractional superstring model-building,
among them the creation of $(1,4)$ models
with chiral fermions and the necessity of compactifying
or interpreting the $(1,4)$ models in four spacetime dimensions.
As we will see, these issues are intimately connected, and
we expect the dictionaries we derive in Sect.~IV to be easily
generalizable to these cases as well.
In Appendix~A we gather together various definitions and properties
of the parafermion characters (or string functions)
which play an important role
throughout our work, and in Appendix~B
we prove an assertion made in Sect.~V.

\endpage

\chapter{FRACTIONAL SUPERSTRINGS}

In this section we provide a self-contained introduction
to the fractional superstring theory as a natural generalization of
the traditional superstring and heterotic string theories.
We also review, where necessary, some relevant features of the
underlying $\bZ_K$ parafermion theories, originally constructed by
Zamolodchikov and Fateev.\refmark\Zam\

As outlined in Sect.~I, the basic idea behind the fractional superstring
is to modify the worldsheet symmetry in such a manner as to obtain
a correspondingly smaller critical spacetime dimension.  In order to do this,
let us begin by
considering the general $SU(2)_K$ WZW
theory.\refmark{\Knizh}\
As is well-known,
this theory consists of primary fields $\Phi^j_m(z)$ which can
be organized into $SU(2)$ representations labelled by an integer
$j$, where $0\leq j\leq K/2$ and $|m|\leq j$ with $j-m\in {\bf Z}$
(for simplicity we are considering only the
holomorphic components).  Since $SU(2)$ always
has a $U(1)$ subgroup which can be bosonized as a free boson
$\varphi$ on a circle of radius $\sqrt{K}$,
we can correspondingly factor these
primary fields $\Phi^j_m$:
$$   \Phi^j_m(z) ~=~ \phi^j_m(z) \,
       \exp\left\lbrace {m\over {\sqrt{K}}}\,\varphi(z) \right\rbrace ~.
\eqn\primariesfactored
$$
Here $\varphi$ is the free $U(1)$ boson,
and the $\phi^j_m(z)$ are the primary fields of
the coset $SU(2)_K/U(1)$ theory.  This coset theory
is the well-known $\bZ_K$ parafermion theory,\foot{
   It is important to realize that for fixed $K$, the $\bZ_K$
   parafermion theory can be realized in more than one way.
   Different $\bZ_K$ models will have different field contents,
   where any $[\phi^j_m]$ can appear with multiplicity other
   than one.  In general, the coupling constants will be different
   as well.}
and these fields $\phi^j_m$ are the corresponding parafermion
fields with highest weights (or conformal dimensions):
$$  h^j_m ~=~ {{j(j+1)}\over{K+2}} ~-~ {{m^2}\over K}~~~~~~~~~~
    {\rm for}~|m|\leq j~.
\eqn\highestweights
$$
The fusion rules
of these parafermion fields $\phi^j_m$ follow from those of
the $SU(2)_K$ theory:
$$  [\phi^{j_1}_{m_1} ] \,\otimes \,
  [\phi^{j_2}_{m_2} ] ~=~ \sum_{j=|j_1-j_2|}^r \, [\phi^j_{m_1+m_2}]
\eqn\fusionrules
$$
where $r\equiv {\rm min}(j_1+j_2,K-j_1-j_2)$
and where the sectors $[\phi^j_m]$ include the
primary fields $\phi^j_m$ and their descendants.
The characters for $[\phi^j_m]$ are $\eta c^{2j}_{2m}$, where
$\eta$ is the Dedekind $\eta$-function and the $c^\ell_n$ are
the so-called string functions;\refmark{\kacpeterson,\Gep}\
these functions will be discussed in more detail in Sect.~III, and
definitions and properties of these functions are
collected in Appendix~A.

Upon factorizing the primary fields $\Phi^j_m$ as in \primariesfactored,
one finds that the $SU(2)$ currents factorize as well:
$$   \eqalign{
    J^+ ~&=~ \sqrt{K} \, \psi_1\,e^{i\varphi/\sqrt{K}}\cr
    J^0 ~&=~ \sqrt{K/2} ~ i\,\partial\varphi \cr
    J^- ~&=~ \sqrt{K} \, \psi_1^\dagger \,e^{-i\varphi/\sqrt{K}}\cr}
\eqn\sutwocurrents
$$
where the parafermion currents $\psi_i \sim \phi^0_i \sim \phi^{K/2}_{K/2-i}$
and $\psi_i^\dagger\equiv \psi_{K-i}$ have conformal dimensions
$i(K-i)/K$ in accordance
with \highestweights.  From \highestweights\ and \fusionrules,
we see that the parafermion stress-energy tensor
$T_{\rm para}(z)\equiv T_{SU(2)_K}(z)-T_\varphi(z)$ and the parafermion
currents $\psi_i$, $i=1,2,...,K-1$ form a closed algebra, namely,
the $\bZ_K$ parafermion current algebra.

Note that $\psi_1$ acting
on a field $\phi^j_m$ increases the $m$ quantum number by one
but does not change the $SU(2)$ spin $j$.
Specifically, we can perform a mode-expansion for $\psi_1$
$$  \psi_1(z) ~=~ \sum_{n\in{\bf Z}} \,A_{d+n} \,z^{-1-d-n+1/K}
\eqn\modeexpansion
$$
where $d$ is a fractional number and the conformal dimension
of $A_{d+n}$ is $-(d+n)$; the $A^\dagger$ mode-expansion for
the $\psi_1^\dagger$ field can be handled similarly.  In \modeexpansion,
of course, the value of $d$ must be chosen appropriately for the
particular parafermion field on which $\psi_1(z)$ is to operate, \eg,
for consistency we must choose a $\psi_1$ moding with $d=(2m+1)/K$
when operating on $\phi^j_m$.
We thus have
$$  \eqalign{
             A_{(2m+1)/K+n}:&~~~~~[\phi^j_m]~\to~[\phi^j_{m+1}]\cr
     A^\dagger_{(1-2m)/K-n}:&~~~~~[\phi^j_m]~\to~[\phi^j_{m-1}]~\cr }
\eqn\mshift
$$
where $n$ are integers.
There is also another special field in the parafermion theory,
namely the energy operator $\epsilon\equiv
\phi^1_0$;
operating on a field $\phi^j_m$ with $\epsilon$ preserves
the $m$ quantum number but yields sectors with $j$ quantum numbers $j+1$,
$j$, and $j-1$.  Specifically, performing a mode-expansion for $\epsilon$
as in \modeexpansion\ and choosing $d$ as indicated below, we find
the actions of its modes:
$$  \eqalign{
     \epsilon_{-2(j+1)/(K+2)+n}:&~~~~~[\phi^j_m]~\to~[\phi^{j+1}_m]\cr
     \epsilon_{n}:&~~~~~[\phi^j_m]~\to~[\phi^{j}_m]\cr
     \epsilon_{2j/(K+2)+n}:&~~~~~[\phi^j_m]~\to~[\phi^{j-1}_m]~.\cr }
\eqn\jshift
$$
Of course, \mshift\ and \jshift\ are valid only when permitted
by the fusion rule \fusionrules.
We thus see that the two fields $\psi_1=\phi^0_1$ and $\epsilon=\phi^1_0$
together allow us to span the entire set of fields $[\phi^j_m]$,
starting from any one of them.

The next step in the fractional superstring construction is
to completely decompactify the free boson $\varphi$, replacing
it with a bosonic field $X$ on a circle of infinite radius;
this field $X$ will be interpreted as corresponding to
a spacetime coordinate.
This decompactification procedure, of course, destroys the underlying
$SU(2)$ WZW symmetry of our theory, but the conformal
invariance remains.

In fact, the symmetry remaining after the boson decompactification
is {\it larger} than simple conformal symmetry.
We can see this for general $K\geq 2$ in the following way.
Let us construct the current\refmark{\Bag}\
$$    \hat J ~\equiv~ \epsilon \,\partial X ~+~ :\epsilon\epsilon:
\eqn\currentJhat
$$
where $\epsilon(z)$ is the energy-operator field
and where $:\epsilon\epsilon:$ is a normal-ordered
product (and is in fact a parafermion descendent of $\epsilon$,
though a Virasoro primary).
For $K\geq 2$, of course, the field $\epsilon$ has conformal dimension
$$  \Delta_\epsilon ~=~ {{2}\over{K+2}} ~,
\eqn\epsilonweight
$$
and it can be shown that the normal-ordered
term $:\epsilon\epsilon:$ has conformal dimension $\Delta_{:\epsilon\epsilon:}
=1+\Delta_\epsilon$.  This is a non-trivial fact, implying
that the normal-ordered field $:\epsilon\epsilon:$ appears in
the $\epsilon\epsilon$ operator product expansion as follows:
$$   \epsilon(z)\epsilon(w) ~=~ {1\over{(z-w)^{2\Delta_\epsilon}}} ~+~
     ...~+~ (z-w)^{1-\Delta_\epsilon}\,:\epsilon\epsilon:(w) ~+~...
\eqn\epsepsOPE
$$
Thus, for $K\geq 2$
the current $\hat J$ in \currentJhat\ has conformal dimension
(or equivalently, spin):
$$   \Delta_{\hat J} ~=~ 1 ~+~ {{2}\over{K+2}} ~=~ {{K+4}\over{K+2}} ~.
\eqn\Jhatweight
$$
On the worldsheet this current
$\hat J$ forms a closed algebra\refmark\Jim\
with $T(z)$, where $T(z)\equiv T_X(z)+T_{\rm para}(z)$
is the stress-energy tensor of the decompactified boson field $X$
plus that of the $\bZ_K$ parafermion theory.
Thus we see that $\hat J$ indeed generates an additional worldsheet symmetry
which we refer to as a {\it fractional}\/ worldsheet supersymmetry.
Note that for $K>2$ the dimensions $\Delta_\epsilon$ and $\Delta_{\hat J}$
are not simple half-integers.  Thus, our underlying worldsheet $(\hat J,T)$
algebra is non-local, with Riemann {\it cuts}\/ (rather than poles) appearing
in the various OPEs.
Note that this $(\hat J,T)$ algebra is merely the simplest
algebra that can be constructed.  For other fractional-superstring
applications, this $(\hat J,T)$ algebra can indeed be extended
to include additional currents.

In order to achieve a sensible interpretation for a $D$-dimensional
spacetime, we associate the decompactified
bosonic field $X$ with a single spacetime coordinate
and tensor together $D$ copies of the $(X,\phi^j_m)$ (or boson plus
$\bZ_K$ parafermion) theory.  We therefore obtain the fractional supersymmetry
current
$$   J~ =~ \epsilon^\mu \,\partial X_\mu ~+~ :\epsilon^\mu \epsilon_\mu:
\eqn\fractionalsupercurrent
$$
where the Lorentz indices $\mu=0,1,...,D-1$
are to be contracted with the Minkowski metric.
A crucial issue, however, is to determine the critical dimension $D$
for arbitrary $K>2$.  A formal
derivation would proceed through the established path:
since the central charge of each $(X,\phi^j_m)$ theory
is
$$      c_0 ~=~ 1 ~+~ {{2K-2}\over{K+2}} ~=~ {{3K}\over {K+2}}~,
\eqn\czero
$$
one would need only to determine the central charge contributed by
the {\it frac\-tion\-ally-su\-per}\/re\-para\-met\-ri\-za\-tion ghosts, and
then choose $D$ so that $D c_0$ exactly cancels this quantity.
Such a calculation is quite difficult, however, and
at present remains to be
done.  Another approach is to examine the Fock space of this
theory and find a dimension $D$ and a corresponding intercept
$-v$ for which extra null states appear;  such work is currently
in progress.
Instead, one can take a third approach\refmark{\letter}\ and
require that any fractional superstring model
produced have a sensible phenomenology --
\eg, a spectrum containing a massless graviton in the case of a closed
string theory, or equivalently a massless vector particle in the case of
an open string theory.
We emphasize that this is a {\it requirement} and not an assumption,
for we are interested in constructing only those fractional string theories
which contain gravity;  other possibilities
are, from this standpoint, phenomenologically unappealing.
It turns out that this requirement is not difficult to implement.

For a general closed string theory, the graviton and the massless
vector both arise from the same right-moving excitation state $|V\rangle_R$;
they differ only in that they are tensored with dissimilar left-moving
states $|V^\prime\rangle_L$.  Therefore, our requirement simply becomes
a requirement on the state $|V\rangle_R$ -- we must demand that
this state exist (\ie, satisfy the physical-state conditions) and
be massless.
For a graviton or spacetime vector particle in a string theory
with arbitrary $K\geq 2$, this state is
$$   \ket{V}_R ~=~ \zeta_\mu\,\epsilon^\mu_{-2/(K+2)}\,\ket{p}_R
\eqn\rightket
$$
where $\ket{p}_R$ is the right-moving vacuum state with momentum $p$,
$\zeta_\mu$ is a polarization vector, and where $\epsilon^\mu_{-2/(K+2)}$
is the lowest excitation mode of the parafermion field $\epsilon$.
Note that the moding of the $\epsilon$ field follows
from its conformal dimension.
This state is indeed the analogue of what appears for the $K=2$ case,
in which a single lowest-mode
excitation of a worldsheet Neveu-Schwarz fermion produces the needed
state:  $\ket{V}_R =\zeta_\mu \epsilon^{\mu}_{-1/2}
\ket{p}$ where $\ket{p}$
is the usual Neveu-Schwarz vacuum state
and $\epsilon_{-1/2}^{\mu} =b_{-1/2}^{\mu}$
is the lowest Neveu-Schwarz creation operator.
(For the bosonic string the $\epsilon$ field is absent, and
we accordingly substitute the bosonic
creation operator $a_{-1}^{\mu}$ for
$b^{\mu}_{-1/2}$ or $\epsilon_{-2/(K+2)}^\mu$.
The same argument then applies.)
Thus, requiring the state \rightket\ to
be massless, we find in general that the vacuum state $\ket{p}$ in \rightket\
must have vacuum energy
${\rm VE}= -v= -2/(K+2)$.   This information can also be stated
in terms of the fractional superstring
character $\chi(q)$.  In general the character has
a $q$-expansion of the form
$$  \chi(q)~=~ \sum_n\, a_n \,q^n
\eqn\charform
$$
where $q\equiv \exp(2\pi i \tau)$, $\tau$ is the complex modular
parameter of the torus, and where $a_n$ is the
number of propagating degrees of freedom at mass level $M^2=n$.
We thus see that the fractional superstring characters must take the form
$$  \chi(q)~=~ q^{-v}\,(1+...) ~=~ q^{-2/(K+2)}\,(1+...)
\eqn\characterone
$$
where inside the parentheses all $q$-powers are non-negative integers.

The only remaining non-trivial physical-state condition on the
state $\ket{V}_R$ is
$$     J_{2/(K+2)}\,\ket{V}_R ~=~0
\eqn\physicalstate
$$
where $J_n$ are the modes of the fractional supercurrent
\fractionalsupercurrent;
this constraint can indeed be shown\refmark{\letter}\ to
yield the expected transversality
constraint $p\cdot \zeta=0$ on the polarization vector $\zeta_\mu$
(which is consistent with its interpretation as the polarization
vector of the massless vector state $\ket{V}_R$).
The state $\ket{V}_R$ therefore has only $D-2$ polarizations (or degrees of
freedom), and from this it follows that at all mass levels of
the physical spectrum
only $D-2$ transverse dimensions worth of polarization states are
propagating degrees of freedom.
Note that this latter assertion is {\it not}\/ an additional
assumption, but rather follows
directly from modular invariance.
We can see this as follows.
Removal of the longitudinal and time-like
components of the massless state $\ket{V}_R$ implies
the removal of the corresponding $q^0$ terms in the products
of string functions which (as we will see in Sect.~III)
comprise the total fractional-superstring partition functions.
However, the string functions $c^\ell_n$, closing as they do under $S$ and
$T$ modular transformations, form an admissible representation of
the modular group.
Modular invariance therefore requires the
removal of two powers of string functions
from the total fractional-superstring partition functions,
which in turn implies that there is indeed a large string gauge symmetry
whose gauge-fixing provides the physical conditions
effectively striking out
all states involving longitudinal and/or time-like
modes from every mass level in the theory.

However, we now recall that conformal invariance requires the
character $\chi(q)$ to take the form
$$   \chi(q)~=~ q^{-c/24}\,(1+...)
\eqn\charactertwo
$$
where $c$ is the total effective conformal anomaly of the propagating
degrees of freedom.  Since each spacetime dimension contributes
the central charge $c_0$ given in \czero, and since we have
determined that effectively only $D-2$ dimensions contribute
to propagating fields, we have
$c=(D-2)c_0$.  Thus, comparing \characterone\ and \charactertwo\
and substituting \czero, we find the result
$$  D~=~ 2 ~+~ {{16}\over K} ~,~~~~~~~K \geq 2~.
\eqn\Dcrit
$$
We see, then, that the critical dimension of the fractional superstring
is a function of the level $K$ of the $SU(2)_K/U(1)$ coset
(\ie, of the $\bZ_K$ parafermionic theory):
for $K=2$ we have $D=10$, for $K=4$ we have $D=6$,
and for $K=8$ we obtain $D=4$.
For $K=1$ we need only set $v=1$
in \characterone\ [as explained after \rightket], whereupon
the argument above yields $D=26$.
Thus, by appropriately
choosing $K$ and building the corresponding worldsheet fractional
supersymmetry as discussed above, we can hope to obtain a series of new
string theories with a variety of critical dimensions.

It is easy to check that the $K=1$ and $K=2$ cases
correspond to the bosonic string and superstring respectively.
For $K=1$, we know that $SU(2)_1$ can be bosonized with
a single boson $\varphi$;  therefore, the $SU(2)_1/U(1)$ coset
(or the $\bZ_1$ ``parafermion'' theory) is trivial,
containing only the identity field $\phi^0_0$.
There are therefore no partners for our bosonic fields $\varphi$
(or the decompactified fields $X^\mu$ which they become),
and hence for $K=1$ the only worldsheet symmetry is conformal symmetry.
This reproduces, of course, the worldsheet structure of the
bosonic string.
Similarly, for $K=2$, the coset $SU(2)_2/U(1)$ (\ie,
the $\bZ_2$ parafermion theory) is simply the
Ising model;  the field content consists of $\phi^0_0$
(the identity), $\phi^1_0 =\phi^0_1$ (which is
of course the $\epsilon$ field, or equivalently
the parafermion current $\psi_1$, or equivalently the Majorana fermion $\psi$),
$\phi^{1/2}_{1/2}$ (the spin field $\sigma$), and $\phi^{1/2}_{-1/2}$
(the conjugate spin field $\sigma^\dagger$).
However, excitations of the worldsheet spin fields
are responsible for spacetime fermions, and indeed together
these fields form a Ramond worldsheet fermion.  The same is
true for the identity and $\psi$ fields:  when appropriately
mixed they form a Neveu-Schwarz worldsheet field whose
excitations ultimately produce spacetime bosons.
Thus, we see that the field content of the $SU(2)_2/U(1)$
theory is simply that of a free worldsheet Majorana fermion theory,
and the symmetry relating this theory to the bosonic
$X^\mu$ fields is an ordinary worldsheet supersymmetry.  This
then reproduces the traditional superstring, and
the heterotic string is the left/right tensoring of
a $K=1$ and $K=2$ theory respectively.
It is therefore apparent that this fractional superstring
language provides a natural framework in which to classify
and uniformly handle all of the traditional string theories,
and in so doing it also points the way
to their non-trivial generalizations.

\endpage

\chapter{PARTITION FUNCTIONS FOR FRACTIONAL SUPERSTRINGS}

Having thus presented the underlying basis for the fractional
superstring theories, we turn our attention to the partition functions
that these theories must have.
We begin by reviewing the generic forms of these partition functions,
and start with the known $K=1$ and $K=2$ cases to establish
our notation and conventions.
We will see, once again, that the $K>2$ string theories
have partition functions whose forms are straightforward extensions of
those of the traditional cases.  This will therefore permit us to
write the partition functions for all of the $(K_L,K_R)$ theories
we shall consider in a common language.

\section{Traditional String Theories}

The first case to consider, of course, is the pure bosonic string;  since
there is no left-moving or right-moving worldsheet supersymmetry,
we may refer to this in our new fractional superstring language
as the $(K_L,K_R)=(1,1)$ case.
As discussed in Sect.~II, the critical spacetime dimension
for this string theory is $D=26$, and thus
this theory contains $26$ bosonic worldsheet fields $X^\mu$, each
of which contributes
to the total one-loop partition function a factor
$$  {\rm each~boson}~\Longrightarrow~ {1\over{ \sqrt{\tautwo} \,
     \eta\overline{\eta} }} ~.
\eqn\bosonfactor
$$
Here $\eta(\tau)$ is the well-known Dedekind $\eta$-function,
and $\tau\equiv \tauone  + i \tautwo$ is the torus modular parameter.
Note that this factor is explicitly real because the left- and right-moving
components of each boson contribute equally.
Since we have already seen that only $D-2=24$ transverse
degrees of freedom propagate
(as is evident in a light-cone gauge approach),
our total partition function for the $(1,1)$ string
is therefore of the form
$$   \calZ_{(1,1)} ~=~ \tautwo^{-12}  \, {1\over{|\eta|^{48}}} ~=~
       \tautwo^{-12} |\Delta|^{-2}~
\eqn\oneonestring
$$
where we have defined $\Delta\equiv \eta^{24}$.
Note that the power of the $\tautwo$ prefactor, which we will
denote $k$, is in general given by
$$   k~=~ 1~-~D/2~
\eqn\kDformula
$$
where $D$ is the number of spacetime dimensions in which the
theory is formulated.  Hence, for the $(1,1)$ string in $D=26$
dimensions, we find $k=-12$, in agreement with \oneonestring.
In the theory of modular functions this quantity $k$ is
known as the modular {\it weight};  the $\eta$-function transforms
under $S:\tau\to 1/\tau$ and $T:\tau\to\tau+1$ as a modular
function of weight $\half$, and therefore in \oneonestring\
the holomorphic and anti-holomorphic pieces separately have
modular weight $k= -12$.  Modular-invariance for the entire
partition function $\calZ$ requires that the modular weight
of its holomorphic factors equal that of its anti-holomorphic
factors, and that this weight also equal the power of the overall
$\tautwo$ factor.
We can see that in \oneonestring\ this is indeed the case.

The next string to consider
is the traditional {\it super}\/string:  since this theory
has a full superconformal worldsheet symmetry for
both the left- and right-moving sectors, in our new language
this is simply the $(K_L,K_R)=(2,2)$ string.
That its critical dimension is 10
follows from \Dcrit, or equivalently from the traditional
argument in which the conformal anomaly $c= -26+11 = -15$ from
the reparametrization ghosts and their superpartners must
be cancelled by the
contributions from $D$ worldsheet boson/fermion pairs,
each of which has total central
charge $c_0=1+\half={\textstyle {3\over 2}}$
[in accordance with \czero\ for $K=2$].
In light-cone gauge only the eight transverse field pairs propagate, and
therefore this theory has a total worldsheet field content consisting
of eight real bosons
(each with a left-moving and right-moving component)
and eight Majorana fermions (or eight Majorana-Weyl left-movers
and eight Majorana-Weyl right-movers).
These fermions, we recall, can be described in our new
viewpoint as simply the fields of the $SU(2)_2/U(1)$ Ising model,
and we are free to group pairs of these Majorana-Weyl fermions
together to form single complex Weyl fermions.
The factor contributed by each of the eight bosons to the
one-loop partition function is given in \bosonfactor,
and the contribution of each complex Weyl fermion is
$$     {\rm each~Weyl~fermion}~\Longrightarrow~
        {{\vartheta}\over{\eta}} ~
\eqn\fermionfactor
$$
where $\vartheta$ represents one of the well-known Jacobi
$\vartheta$-functions defined in
Appendix~A.\foot{Which particular $\vartheta$-function
   is appropriate depends
   on the boundary conditions assigned to the fermion as it traverses the
   two cycles of the torus.  Only periodic or antiperiodic boundary
   conditions will yield the Jacobi $\vartheta$-functions, but if
   a fermion is chosen to have periodic/periodic boundary conditions then
   its zero-modes cause the total partition function to vanish identically.}
Since the contributions of right-moving fields are the
complex-conjugates of those of left-moving fields,
the total partition function for the $(2,2)$ string
therefore takes the form
$$   \calZ_{(2,2)}~\sim~ \tautwo^{-4} \,{1\over{|\eta|^{16}}}\,
    \sum\,\left| {{\vartheta}\over{\eta}}\right|^8 ~\sim~
     \tautwo^{-4} \, \Delta^{-1/2} \,\deltabarminushalf
             \,\sum\,\vartheta^4  \,\varthetabar^4 ~,
\eqn\twotwoform
$$
where $\vartheta^4$ indicates four (not necessarily identical)
Jacobi $\vartheta$-function factors.
The summation in \twotwoform\ is over fermion boundary conditions,
as is needed in order to achieve modular invariance.
In this section
we shall frequently use the schematic notation in \twotwoform,
employing the relation $\sim$
when indicating the general forms of partition functions.
Note that since each Jacobi $\vartheta$-function transforms under
the modular group with weight $1/2$, we see that
the holomorphic and anti-holomorphic factors in \twotwoform\ --- as well
as the $\tautwo$ power --- imply that $k= -4$ for this theory.
This result is once again in agreement with
our spacetime dimension $D=10$ according to \kDformula.

We shall be most concerned in this paper with heterotic-type
string theories.  The traditional heterotic string has a right-moving
worldsheet supersymmetry as in the superstring;  hence the right-moving
propagating field content must consist, as before, of eight bosons
and eight Majorana-Weyl fermions, with critical spacetime dimension
$D=10$.  As above, this is simply the choice
$K_R=2$ for the right-moving sector.
However, the left-moving sector of the theory
must also have the eight transverse left-moving components of our bosonic
fields, and thus in order to cancel the $c=26$ contribution from
the left-moving reparametrization ghosts we must additionally have
32 Majorana-Weyl fermions (or equivalently 16 complex Weyl fermions,
or 16 compactified scalar bosons $\phi$).  Our traditional heterotic
string is therefore a $(K_L,K_R)=(1,2)$ theory with
partition function of the form
$$ D=10:~~~~~ {\cal Z}_{(1,2)}~\sim~ \tautwo^{-4}
      \,\deltaminusone \,\deltabarminushalf
    \,\sum\, \vartheta^{16} \,\varthetabar^{4}~.
\eqn\onetwotheorytendim
$$
Note that we continue to have $k= -4$ from each factor in
this partition function, in agreement with \kDformula.
As examples of \onetwotheorytendim\ which will be relevant later,
we can look at those known $(1,2)$ $D=10$ theories (or models) which
have spacetime supersymmetry.
As is well-known,\refmark{\heteroticrefs}\
there are only two such self-consistent models:  these have gauge
groups $SO(32)$ and $E_8\otimes E_8$.  Their partition
functions respectively are as follows:
$$  \eqalign{
    D=10,~SO(32):&~~~~~
    \calZ ~=~ (\half)^2 \, \tautwo^{-4} \, \calK\,
     (\beta^4 + \gamma^4 + \delta^4) \cr
    D=10,~E_8\otimes E_8:&~~~~~
    \calZ ~=~ (\half)^3 \, \tautwo^{-4} \, \calK\,
     (\beta^2+\gamma^2+\delta^2)^2 \cr
   }
\eqn\Dequaltenmodels
$$
where we have made the following definitions which we will use
throughout:
$$   \eqalign{
 \beta ~\equiv&~\thetatwo^4~,
   ~~~\gamma~\equiv~\thetathree^4~,~~~
      \delta~\equiv~\thetafour^4~,
   \cr
 J~\equiv&~\gamma-\beta-\delta~,\cr
 \calK~\equiv&~ \deltaminusone \,\deltabarminushalf \,\Jbar ~.\cr}
\eqn\definitions
$$
Note that the partition functions \Dequaltenmodels\ are indeed of
the form \onetwotheorytendim, with the holomorphic parts factorizing
into group characters to reflect the underlying group structure.
The quantity $J$ in \definitions\
is of course the Jacobi factor, and the spacetime supersymmetry
of these models
(or equivalently, the vanishing of their partition functions)
arises from the identity $J=0$.  It is in fact a further identity
that
$$   \beta^4+\gamma^4+\delta^4 ~=~ \half\,(\beta^2+\gamma^2+\delta^2)^2~,
\eqn\latticeidentity
$$
as a consequence of which the two partition functions in
\Dequaltenmodels\ are equal even without use of the Jacobi identity
$J=0$.

It is straightforward to check the modular invariance of these
partition functions \Dequaltenmodels.  For any
$A \equiv {a~b\choose c~d}$ in the modular group
we can define the so-called ``stroke'' operator $[A]:~f\to f[A]$ where
$$  (f[A])(\tau)~\equiv~ (c\tau +d)^{-k}
    \,f\left( {{a\tau+b}\over{c\tau+d}} \right)
\eqn\strokedef
$$
(here $k$ is the modular weight of $f$).  It then follows from
the modular transformation properties of the $\eta$ and $\vartheta$
functions (see Appendix~A) that
$$  \eqalign{
    \beta[S]~=~ -\delta ~,&~~~\beta[T]~=~-\beta\cr
    \gamma[S]~=~ -\gamma ~,&~~~\gamma[T]~=~+\delta\cr
    \delta[S]~=~ -\beta ~,&~~~\delta[T]~=~+\gamma\cr
    \calK [S]~=~ +\calK ~,&~~~\calK [T]~=~+\calK ~.\cr }
\eqn\modtransrules
$$
Thus, $\calK$ is itself modular-invariant
(\ie, invariant under $[S]$ and $[T]$),
and in each partition function the other factors involving $\beta$, $\gamma$,
and $\delta$ must be (and are) themselves modular-invariant as well.

It is also possible to construct self-consistent $(1,2)$ theories
in dimensions $D<10$;  one needs simply, for example, to
re-interpret the extra $10-D$ bosonic degrees of freedom
or in some other manner hide them from low-energy physics
(\eg, through compactification).  We stress, however, that
all such methods do not alter the underlying critical
dimension away from 10;  these theories remain theories of the $(1,2)$
variety.
For example, one method of constructing $(1,2)$ models in
$D=4$ involves fermionization:
each of the six extra left- or right-moving bosonic
degrees of freedom can, for example, be represented in terms
of two free worldsheet Majorana-Weyl fermions (or
a single free Weyl fermion).\REF\KLT{H.~Kawai,
D.C.~Lewellen, and S.-H.H.~Tye, {\it Nucl.~Phys.} {\bf B288}, 1
(1987).}\refend\
The propagating field content for this four-dimensional $(1,2)$ theory
therefore consists of two left-moving and two right-moving bosons,
as well as 22 left-moving and 10 right-moving Weyl fermions.
The total partition function for such a $D=4$ $(1,2)$ theory
hence takes the form
$$ D=4:~~~~~ {\cal Z}_{(1,2)}  ~\sim~
      \tautwo^{-1} \,\deltaminusone \,\deltabarminushalf\,\sum\,
   \vartheta^{22}\, \varthetabar^{10} ~.
\eqn\onetwotheoryfourdim
$$
Note that once again \kDformula\ is satisfied, with $k=-1$ for $D=4$.
As an example, we present the partition function of what is possibly
the simplest such spacetime-supersymmetric model that can be constructed.
Named Model $M1$ in Ref.~[\KLT], it has gauge group $SO(44)$,
the largest possible in $D=4$;  its (modular-invariant)
partition function is
$$ D=4,~SO(44):~~~ \calZ ~=~ (\half)^2 \,\tautwo^{-1}\,
      \calK\, \biggl(|\beta|\bbar \beta^5+
     |\gamma| \cbar \gamma^5 + |\delta|\dbar \delta^5 \biggr)~.
\eqn\fourdimsofourfour
$$
Since the ``maximal'' gauge group allowed in $D$ dimensions is
$SO(52-2D)$, we see that this model is the $D=4$ analogue
of the $D=10$ $SO(32)$ model in \Dequaltenmodels.

We note for future reference that this idea can in fact be generalized
to obtain $(1,2)$-type theories in any spacetime dimension $D\leq 10$.
We in general obtain the partition function form
$$ {\rm general~}D:  ~~~~~
    {\cal Z}_{(1,2)} ~\sim~ \tautwo^k \,\deltaminusone \,\deltabarminushalf\,
      \sum\,\vartheta^n\,\varthetabar^{\overline{n}}~
\eqn\onetwotheoryarbdim
$$
where $k$ is given by \kDformula\ and
$$  \eqalign{
    n~&=~ 26~-~D~=~ 24~+~ 2k ~,\cr
    \overline{n} ~&=~ 14~-~D~=~ 12~+~ 2k ~.\cr  }
\eqn\powers
$$
Much of our work will be concerned with models in $D=6$:  in this
case we have $k= -2$, $n=20$, and $\overline{n}= 8$.  The
``maximally symmetric'' model with spacetime SUSY has gauge group $SO(40)$;
its partition function, along with those of other supersymmetric models
we will be discussing, is as follows:
$$  \eqalign{
  D=6,~SO(40):&~~~
      \calZ~=~(\half)^2 \,\tautwo^{-2}\,\calK\,
       (\bbar \beta^5+\cbar \gamma^5+\dbar \delta^5)\cr
  D=6,~SO(24)\otimes E_8:&~~~
      \calZ~=~(\half)^3 \,\tautwo^{-2}\,\calK\,\times\cr
   &~~~~~~~~~~~~~~~~~\times\,
       (\bbar \beta^3+\cbar \gamma^3+\dbar \delta^3)\,
     (\beta^2+\gamma^2+\delta^2)\cr
  D=6,~SO(24)\otimes SO(16):&~~~
      \calZ~=~(\half)^3 \,\tautwo^{-2}\,\calK
        ~ \biggl\lbrace \bbar \beta^2(\beta^3+\gamma^3-\delta^3) ~+\cr
 ~&~~~~~~+~\cbar \gamma^2(\beta^3+\gamma^3+\delta^3)~+~
         \dbar \delta^2(-\beta^3+\gamma^3+\delta^3)\biggr\rbrace~.\cr}
\eqn\Dequalsixmodels
$$
We again observe the familiar factorization, with the same $E_8$ factor
above as in \Dequaltenmodels.  It is simple, using \modtransrules,
to see that these partition functions are indeed modular-invariant.

\section{Fractional Superstring Theories}

The partition functions for {\it fractional}\/ superstrings can
be determined in precisely the same manner as above.
Since we know the critical dimension $D$ for a given value of $K$,
we can readily deduce the forms that partition functions must take
for general $(K_L,K_R)$ combinations.
The factor contributed to the total partition function from
each worldsheet boson is given, as before, by \bosonfactor;
recall that this factor includes the contributions from both holomorphic
and anti-holomorphic (or left- and right-moving) components.
The factor contributed by each worldsheet parafermion, however, is a
generalization of \fermionfactor;  in general we have
$$  {\rm each~parafermion}~\Longrightarrow~ \eta \, c~
\eqn\parafermionfactor
$$
(the above is for left-moving parafermions;  right-moving
parafermions contribute the complex-conjugate).
Here $\eta$ is the usual Dedekind function, and
$c$ schematically represents one of the so-called
parafermionic {\it string functions}.\refmark{\Gep,\kacpeterson}\
These functions are defined in Appendix~A, but for our
present purposes we need record only the following facts.
These functions $c^\ell_n$ may be defined in terms of $q$-expansions
which depend on $K$ as well as the two parameters $\ell$ and $n$,
and one string function $c^{2j}_{2m}$ may correspond to more
than one parafermion field $\phi^j_m$ (for example, the distinct
fields $\phi^j_m$ and $\phi^j_{-m}$ both give rise to $c^{2j}_{2m}$).
Each $c^\ell_n$ is an eigenfunction under $T:\tau\to\tau+1$,
and under $S:\tau\to -1/\tau$ they mix forming a closed set;
furthermore, they transform under the $S$ and $T$ transformations with
the {\it negative} modular weight $k= -\half$.

As expected, these string functions (which depend on $K$) should
reduce to the traditional modular functions for the $K=1$ and
$K=2$ cases, and this is indeed precisely what occurs.
For the $K=1$ case there is only one string function $c^0_0$,
and since our $SU(2)_1/U(1)$ theory consists of only
the identity field $\phi^0_0$, we quickly have
$$   K=1:~~~~ \eta\,c^0_0~=~1~~~~\Longrightarrow~~~~
              c^0_0 ~=~\eta^{-1}~.
\eqn\cforKequalone
$$
Thus, for $K=1$ the one string function $c^0_0$ is related
to the boson character $\eta$.
Similarly, for $K=2$ there are precisely three
string functions $c^0_0$, $c^1_1$, and $c^2_0$, and
these are related to the
three fermionic Jacobi $\vartheta$-functions as follows:
$$   K~=~2:~~~~~~~\cases{
     ~2\,(c^1_1)^2 ~&$=~ \thetatwo/\eta^3$\cr
     ~(c^0_0+c^2_0)^2 ~&$=~ \thetathree/\eta^3$\cr
     ~(c^0_0-c^2_0)^2 ~&$=~ \thetafour/\eta^3~$.\cr}
\eqn\cforKequaltwo
$$
For $K>2$, of course, the string functions involve
more than just these simpler functions.

Therefore, looking first at the $(K_L,K_R)=(K,K)$ theories
with $K>2$ (these are the generalizations of the usual Type~II theories),
we see that our worldsheet field content consists (in light-cone
gauge) of $D-2=16/K$ coordinate bosons and $16/K$ each
of left- and right-moving parafermions.  We therefore
obtain the form for the total $(K,K)$ partition function:
$$   \calZ_{(K,K)} ~\sim~
     \tautwo^{-8/K} ~\sum\, {\overline{c}}^{16/K}\,c^{16/K}~.
\eqn\kktheory
$$
Note that $k\equiv -8/K=1-D/2$
in accordance with \kDformula\ and \Dcrit, and
that the $\eta$-functions have
cancelled between the bosonic and parafermionic contributions.
Thus, in our cases of interest we obtain:
$$  \eqalign{
 D=6:~~~~~    \calZ_{(4,4)} ~&\sim~
              \tautwo^{-2} ~\sum\,{\overline{c}}^4 \,c^4\cr
 D=4:~~~~~    \calZ_{(8,8)} ~&\sim~
              \tautwo^{-1} ~\sum\,{\overline{c}}^2 \,c^2\cr }
\eqn\typeiicases
$$

Similarly, one can consider the analogues of heterotic
string theories;  these would be the $(K_L,K_R)=(1,K)$ situations.
The right-moving sectors of these theories are precisely as in the $(K,K)$
cases, and hence the anti-holomorphic parts of their partition functions
have the same form as in \kktheory.
Since their left-moving sectors must already contain the left-moving
components of the coordinate bosons,
we must augment their left-moving field contents to achieve
conformal anomaly cancellation.  The contribution of
the bosons to the central charge is of course $2+16/K$, and
since this sector is a bosonic theory we must cancel
the usual ghost contribution $c= -26$.  This
requires additional matter fields with central charge
$$ \Delta c ~=~ 8 \left( {{3K-2}\over{K}} \right)~,
\eqn\cadd
$$
and we may choose these fields to be $\Delta c$ complex Weyl
worldsheet fermions [each of which has the partition-function
contribution given in \fermionfactor].  Thus, for a general
heterotic-type $(1,K)$ theory we expect a partition function
of the form
$$ \eqalign{
 \calZ_{(1,K)} ~&\sim~ \tautwo^{-8/K}~ \sum\,
     {\overline{c}}^{16/K} \,  \left({1\over \eta}\right)^{16/K} \,
         \left( {{\vartheta}\over \eta} \right)^{8(3K-2)/K} \cr
   ~&\sim~ \tautwo^{-8/K} ~\sum\,{\overline{c}}^{16/K} \,\deltaminusone \,
    \vartheta^{8(3K-2)/K}~. \cr}
\eqn\hetform
$$
Note that since $\Delta c = 8(3K-2)/K= n$ [where $n$ is the quantity
defined in \powers], the holomorphic part of this partition function
is precisely the same as we obtained
in \onetwotheoryarbdim\ for general spacetime dimension $D$.
This is of course to be expected;  these left-moving sectors
are both $K=1$ theories.  The crucial difference, however,
is that the heterotic $(1,K)$ theory is {\it in} its critical
dimension for {\it all}\/ values of $K$.

For completeness, we should also note that it {\it is} possible,
just as in the $(1,2)$ case, to consider $(1,K)$ theories
in spacetime dimensions $D<D_c$.  One can, for example, choose
to retain $D$ of the above bosons and replace the remaining $D_c-D$
bosons with as many worldsheet Weyl fermions.
It is clear that from such a sector the partition function
contribution is
$$   \left( {1\over \eta} \right)^{D-2} ~ (\eta c)^{16/K} ~
       \left( {{\vartheta}\over \eta} \right)^{D_c-D} ~=~
        c^{16/K} \, \vartheta^{D_c-D} ~;
\eqn\dlessthandcrit
$$
the first factor is the contribution of the $D-2$ remaining
bosons, the second is that of the original $D_c-2=16/K$ parafermions,
and the third arises from the fermionized $D_c-D$ bosons.
Thus, for example, there are two ways to achieve a $D=4$ theory:
one can consider a $(1,8)$ theory in its critical dimension,
or a $(1,4)$ theory in which two dimensions are in some way
compactified or fermionized.
We shall discuss this latter possibility in Sect.~VI.

In fact, use of these parafermionic string functions
allows us to collect together in a simple way {\it all}\/
of the partition function
forms we have considered for the general
$(K_L,K_R)$ theory in arbitrary dimension $D\leq D_c$.
For a given value of $K_R$, we have considered
the two cases $K_L=1,K_R$;  this includes all traditional string
theories as well as our new ones.  In general we can write simply
$$  \calZ_{(K_L,K_R)} ~=~ \tautwo^k \,\sum_i\,
   L_i^{(K_L)} \,\overline{R_i^{(K_R)}} ~
\eqn\generalpf
$$
where the $L_i^{(K_L)}$ are the contributions from the left-moving sectors
and the $R_i^{(K_R)}$ are from the right-movers.  In the above formula $k$ is
always given by \kDformula\ (where $D$ is the dimension of spacetime,
not the critical dimension).  If we define the critical dimensions
$D_c^R=2+16/K_R$ and $D_c^L=2+16/K_L$ (or 26 if the corresponding
$K=1$), then the general forms for the $L_i$ and $R_i$ can be given as follows:
$$    L_i^{(K_L)}~\sim~ c^{D_c^L-2}\,\vartheta^{D^L_c-D} ~,~~~~~
      R_i^{(K_R)}~\sim~ c^{D_c^R-2}\,\vartheta^{D^R_c-D} ~.
\eqn\tablesubstitute
$$
We can easily check the special cases $K=1$ and $K=2$.  For
a $(1,1)$ theory formulated in arbitrary $D\leq D_c=26$, \cforKequalone\
yields $L^{(1)}=R^{(1)}=\deltaminusone \,\vartheta^{26-D}$:  for $D=D_c$
this therefore reproduces \oneonestring, and for $D<D_c$ this
reproduces the holomorphic
part of \onetwotheoryarbdim\ for the $(1,2)$ theory.
Similarly, for $K_R=2$ in arbitrary dimension $D\leq D_c$,
\cforKequaltwo\ yields $R^{(2)}= \deltaminusone \vartheta^{14-D}$,
in accord with the anti-holomorphic part of \onetwotheoryarbdim.
Thus, \generalpf\ and \tablesubstitute\
are indeed the most general partition function forms for
the traditional as well as the fractional string theories, brought
together in a natural way through our use of the parafermionic string
functions.

In order to construct sensible partition functions having
the above forms, it is first necessary to find suitable combinations
of the string functions which can replace the $c^{D_c-2}$ factors above.
There are several requirements.
First, we want linear combinations which are tachyon-free:
this means that in a $q$-expansion
$$  \sum \,c^{D_c-2} ~=~ \sum_n \, a_n \,q^{n}
\eqn\expansion
$$
we must have $a_n=0$ for all $n<0$.  Second, we also require
that there be a spectrum of massless particles:  this means
that for at least a {\it subset}\/ of terms corresponding to a massless
sector in the linear combination we must have $a_0\not= 0$.
Third, since we would like to use these linear combinations
in building heterotic partition functions, we must demand
that these combinations be invariant under $T^4: \tau\to\tau+4$
(this is because all of the Jacobi $\vartheta$-function expressions
with which we will be dealing share this property).
[In the language of \expansion, this means
that $a_n=0$ for all $n\not\in {\bf Z}/4$.]
Fourth, demanding modular-invariance for our total partition function
amounts to demanding that any set of linear combinations satisfying
the above constraints must also close under $S:\tau\to -1/\tau$.
Finally, for a given $K$, we must of course demand that
our linear combinations each involve $D_c-2$ powers of level-$K$ string
functions:  for $K\geq 2$ this means
that we require $16/K$ string-function factors,
and for $K=1$ we require 24.

It is clear that linear combinations satisfying
all of the above constraints exist for the $K=1$ and $K=2$ cases;
for $K=1$ we have simply:\foot{
      This expression $A_1$ actually has $a_n\not= 0$ for $n<0$,
      in accordance with the known result that the bosonic string
      contains on-shell tachyonic states.}
$$  A_1 ~=~ (c^0_0)^{24} ~=~ \Delta^{-1} ~,
\eqn\AKequalone
$$
and for $K=2$ we can choose:
$$  \eqalign{
   A_2~&=~ 8\,(c^0_0)^7c^2_0 ~-~ 8\,(c^1_1)^8 ~+~ 56\,(c^0_0)^5(c^2_0)^3
       ~+~ 56\,(c^0_0)^3(c^2_0)^5 ~+~ 8\,c^0_0(c^2_0)^7\cr
      &=~ \half \,\deltaminushalf\,J~.\cr}
\eqn\AKequaltwo
$$
We note that in each of these cases
only one linear combination is necessary to achieve closure under $S$
and $T$, and we observe that $A_2=0$ as a consequence of the Jacobi identity
$J=0$.  This is simply a reflection of the {\it spacetime}
supersymmetry of the superstring.
These two expressions have, of course, already been determined:
$\calZ_{(1,1)}$ in \oneonestring\ is merely $\tautwo^{-12}|A_1|^2$,
and $\tautwo^{-4}|A_2|^2$ is merely a special (spacetime-supersymmetric)
case of $\calZ_{(2,2)}$ in \twotwoform.

Similarly, it is possible to construct linear combinations
satisfying all of the above constraints for the
$K=4$ and $K=8$ cases as well.\refmark{\letter}
These are as follows.  For $K=4$ we have the two
combinations:
$$ \eqalign{
  A_4 ~&=~ 4\,(c^0_0~+~c^4_0)^3\,(c^2_0)~-~4\,(c^2_2)^4
   ~+~32\,(c^4_2)^3\,(c^2_2)~-~4\,(c^2_0)^4~,\cr
  B_4 ~&=~ -\,4\,(c^2_2)^2\,(c^2_0)^2
   ~+~8\,(c^0_0~+~c^4_0)\,(c^2_0)(c^4_2)^2 \cr
   &~~~~+~4\,(c^0_0~+~c^4_0)^2\,(c^2_2)(c^4_2)~ \cr}
\eqn\ABKequalfour
$$
and for $K=8$ we have the three combinations:
$$\eqalign{
  A_8 ~&=~ 2\,(c^0_0~+~c^8_0)\,(c^2_0~+~c^6_0)~-~2\,(c^4_4)^2
   ~+~8\,(c^8_4c^6_4)~-~2\,(c^4_0)^2~,\cr
  B_8 ~&=~ 4\,(c^0_0~+~c^8_0)(c^6_4) ~+~ 4\,(c^2_0~+~c^6_0)\,(c^8_4)
   ~-~4\,(c^4_0c^4_4)~,\cr
  C_8 ~&=~ 4\,(c^2_2~+~c^6_2)\,(c^8_2~+~c^8_6)~-~4\,(c^4_2)^2~.\cr }
\eqn\ABCKequaleight
$$
[In fact, combinations satisfying these requirements exist
for the $K=16$ ($D=3$) case
as well.\REF\ADT{P.C.~Argyres, K.R.~Dienes,
    and S.-H.H.~Tye, {\it New Jacobi-like Identities for $\bZ_K$
    Parafermion Characters},
    Cornell/McGill preprint CLNS 91/1113 = McGill/91-37
    (to appear).}\refend]
The transformation properties of these functions under $S$ and
$T$ will be given in Sect.~IV;
in particular,
each of these expressions has a $q$-expansion of the form
$q^h(1+...)$, where inside the parentheses all $q$-exponents
are non-negative integers and where $h=0$ for $A_4$ and $A_8$,
$h=1/2$ for $B_4$ and $B_8$, and $h=3/4$ for $C_8$.
Thus, we see that massless spacetime particles
can contribute to the these partition functions only through
$A_4$ and $A_8$:
indeed, the terms $(c^0_0)^{D_c-3}c^2_0$ can be interpreted
as arising from massless spacetime vector particles, and
the terms $-(c^{K/2}_{K/2})^{D_c-2}$
can be interpreted as corresponding to massless spacetime
fermions obeying the Dirac equation.\refmark{\letter}\
We will, therefore, take these combinations as our basic
building blocks when constructing our fractional superstring
partition functions, substituting them for
the $c^{D_c-2}$ factors in the $K=4$ and
$K=8$ partition functions respectively.

In fact, it turns out that there is one additional
remarkable property shared by the expressions in
\ABKequalfour\ and \ABCKequaleight.  Just as
$A_2$ vanishes due to the Jacobi $\vartheta$-function identity
$J=\gamma-\beta-\delta=0$, it can be shown\refmark{\ADT}\
that each of these new parafermionic
string-function expressions vanishes as well:
$$  A_4~=~B_4~=~A_8~=~B_8~=~C_8~=~0~.
\eqn\paraJacobi
$$
Thus, for $K=4$ we have
the two new Jacobi-like identities $A_4=B_4=0$,
and for $K=8$ we have the three new identities
$A_8=B_8=C_8=0$.  (Similar results exist for the $K=16$ case as
well.\refmark{\ADT})
This result is quite important,
for \paraJacobi\ can then serve as the mechanism by which any
$(K_L,K_R)$ partition function
can be made consistent with spacetime supersymmetry.
Any $(K_L,K_R)$ model with spacetime supersymmetry
must therefore have partition functions built from these
Jacobi-like factors, and the $K=2$ full Jacobi identity
$A_2=0$ appears merely as the $K=2$ special case
of the more general identities for general even integers $K$.

We emphasize that
such $B$ and $C$ sectors, necessitated by closure
under $[S]$ for the $K=4$
and $K=8$ theories, are completely new features
arising only for the $K>2$ fractional superstrings,
and as such they are responsible for much of the new
physics these strings contain.
For example, not only do they lead (as we have seen) to {\it multiple}
independent Jacobi identities, but they may also be responsible
for {\it self-induced}\/ compactifications of these strings.
We will discuss this possibility in Sect.~VI.

In fact, the Jacobi identity is not the only famous identity
which generalizes to higher $K$.
For $K=2$, there is another well-known identity involving the
$\eta$ and $\vartheta$-functions:
$$     \vartheta_2 \,\vartheta_3 \,\vartheta_4 ~=~ 2\,\eta^3~;
\eqn\bosonization
$$
this identity relates the fermion characters $\vartheta_i$
to the boson character $\eta$,
and hence we may refer to this as a bosonization identity.
It turns out\refmark{\ADT}\ that \bosonization\ is only the
$K=2$ special case of another series of identities, each relating
the more general $\bZ_K$ parafermion
characters $c^\ell_n$ for a different $K\geq 2$
to the boson character $\eta$.
In fact, there also
exist various {\it other}
series of identities whose $K=2$ special cases
are known and have well-understood physical interpretations;
proofs of all of these identities, as well as methods
for generating them for arbitrary $K$, are presented in Ref.~[\ADT].
We therefore see that the string functions $c^\ell_n$ provide a uniquely
compelling language in which to generalize previously known string-theory
results, providing us with new Jacobi-like and other identities whose
physical interpretations are only beginning to be explored.
Thus we see again (this time on the partition-function level)
that by leading directly to these string-function expressions and identities,
the fractional superstring construction
does indeed provide us with a natural means of generalizing the traditional
string theories.

\endpage

\chapter{DICTIONARIES FOR MODEL-BUILDING}

Having established that spacetime supersymmetry can
be incorporated by using the expressions \ABKequalfour\
and \ABCKequaleight, we now turn our attention to
the construction of
actual supersymmetric $(K_L,K_R)$ models.
We focus our attention primarily on the heterotic $(K_L,K_R)=(1,4)$
and $(1,8)$ cases, and construct a procedure for
generating models in these classes.
Our procedure involves ``translating'' or drawing
correspondences between the $(1,K)$ models and known
$(1,2)$ models in $D=2+16/K$, and we construct ``dictionaries''
which enable these translations to take place for a given $K$.
We find that these dictionaries are intuitive and practical,
and furthermore (as we will see in Sect.~V)
they yield substantially and understandably smaller
spaces of $(1,K)$ models than $(1,2)$ models in $D<10$.
In particular, we will find that only those $(1,2)$ models which
have a maximal number of spacetime supersymmetries are translatable;
these are the models with $N=N_{\rm max}$ SUSY, where
$$   N_{\rm max}~=~ \cases{
      1 & for $D=10$\cr
      2 & for $D=6$\cr
      4 & for $D=4$~.\cr }
\eqn\nmax
$$

We should point out that
throughout this and the next section we will be focusing
our attention on those $(1,2)$ models whose partition
functions can be built from Jacobi $\vartheta$-functions,
in accordance with the presentation in Sect.~III.
This is indeed a broad class of models, but it is not
all-inclusive.  However, the dictionaries we will be developing for
these models rest on very general principles, and we expect
this dictionary idea to be equally applicable to other methods
of $(1,2)$ model-construction as well.

\section{Modular Invariance}

As discussed in Sect.~III, models of the $(1,K)$
variety in their critical dimensions $D=2+16/K$ must have partition
functions of the form:
$$ \calZ_{(1,K)}~\sim~ \tautwo^k \, \sum\, {\overline{c}}^{16/K}
     \, \deltaminusone \, \vartheta^{24-16/K}~,
\eqno\eq
$$
and spacetime supersymmetry can be incorporated by allowing the factors
${\overline{c}}^{16/K}$ to take the values
$\overline{A_K}$, $\overline{B_K}$, and (for $K=8$) $\overline{C_K}$.
Let us change notation slightly, and refer to these quantities
collectively as $\overline{A^i_{(K)}}$
where $i=1,2,3$ for $K=8$ and $i=1,2$ for $K=4$.  Supersymmetric models
will therefore have partition functions
$$    \calZ_{(1,K)} ~=~ \tautwo^k \,\sum_i \,\overline{A^i_{(K)}}\,F^i_{(K)}
\eqn\fs
$$
where the $F$'s take the forms
$$    F^i_{(4)}  ~\sim~ \deltaminusone \,\sum\,\vartheta^{20}
                    ~~~~~{\rm and}~~~~~~
      F^i_{(8)}  ~\sim~ \deltaminusone \,\sum\,\vartheta^{22}~.
\eqn\fforms
$$

Let us now determine the constraints we must impose on the $F$'s
due to modular-invariance.
It follows from the transformation properties of the individual
string functions $c^\ell_n$
under the $[S]$-transformation (see Appendix~A) that
the expressions $A^i_{(K)}$ transform as
$$  A_{(K)}^i[S]~=~ e^{4\pi i/K} \,\sum_{j}\, M_{(K)}^{ij} \,A_{(K)}^j
\eqn\AtransformationS
$$
where the matrices $M_{(K)}$ are
$$   M_{(4)}~=~ \pmatrix{ 1/2 & 3 \cr 1/4 & -1/2} ~,~~~~
     M_{(8)}~=~ \pmatrix{ 1/2 & 1/2 & 1 \cr
                                       1/2 & 1/2 & -1 \cr
                                       1/2 & -1/2 & 0 \cr} ~,
\eqn\Mmatrices
$$
and where the stroke operator $[S]$ was defined in \strokedef.
Demanding $[S]$-invariance of the total partition function $\calZ$
therefore yields
$$ \calZ[S]
    ~=~ e^{-4\pi i/K}\, \tautwo^{k} \, \sum_{i,j} \, M^{ij}\,\overline{A^j}
       \,F^i[S]     ~\equiv~  \tautwo^{k} \sum_j\,\overline{A^j}\,
     F^j~,
\eqno\eq
$$
or
$$       F^i ~=~ e^{-4\pi i/K}\, \sum_j\, (M^t)^{ij}\,F^j[S]
\eqn\something
$$
where $M^t$ is the transpose of $M$.
Since $M^2=(M^t)^2={\bf 1}$ for both cases $K=4$ and $8$,
we can immediately solve \something\ for $F^j[S]$, yielding the constraint
$$       F^i[S] ~=~ e^{4\pi i/K}\,\sum_j\, (M^t)^{ij}\,F^j~.
\eqn\FconstraintS
$$

Similarly, under $[T]$ the $A$'s have the transformation
$$  A_{(K)}^i[T]~=~ \sum_{j}\, N_{(K)}^{ij} \,A_{(K)}^j
\eqn\AtransformationT
$$
where the matrices $N_{(K)}$ are
$$    N_{(4)}~=~ \pmatrix{ 1 & 0 \cr 0 & -1 \cr}~,~~~~~
      N_{(8)}~=~ \pmatrix{ 1 & 0 & 0 \cr 0 & -1 & 0 \cr
                        0 & 0 & -i\cr }~.
\eqn\Nmatrices
$$
Proceeding precisely as above, we find the additional constraints:
$$       F^i[T]~=~ \sum_j\, N^{ij}\,F^j~,
\eqn\FconstraintT
$$
which in this case imply simply that each $F^i$ must transform
under $[T]$ precisely as does the corresponding $A^i$.
This is also clear
from a $q$-expansion:  invariance under $T:\tau\to\tau+1$ means
that in $\calZ=\tautwo^k \sum_{m,n}a_{mn}\qbar^m q^n$ we can have
$a_{mn}\not= 0$ only if $m-n\in {\bf Z}$.  Since each $A^i$ has a
$q$-expansion of the form $A=q^h(1+...)$ where inside the
parentheses all powers of $q$ are integral, we see that the
corresponding $F^i$ must take the same form with the same
value of $h$.  This is the content of \FconstraintT.

We now give a general procedure for obtaining expressions
$F^i$ of the forms \fforms\ which satisfy both \FconstraintS\
and \FconstraintT.
Let us first examine the $K=8$ case, after which the $K=4$ case
will be straightforward.
In order to do this, we consider the space $\calF$ of polynomials
in the three quantities $\lbrace\thetatwo^2, \thetathree, \thetafour\rbrace$
(so that $f[T^4]=f$ for all $f\in \calF$),
and establish four projection operators $P_\ell$ ($\ell=0,...,3$) in
this space.  These operators $P_\ell$ are defined
$$  P_\ell\,f ~\equiv~ \quarter \,\sum_{n=0}^3 \,
    \exp\left\lbrace - {{i\pi \ell n}\over 4} \right\rbrace\,f[T^n]
\eqn\projections
$$
where $f\in \calF$ is any polynomial in this space;  note that
with this definition $\sum_\ell P_\ell={\bf 1}$ and $P_\ell P_{\ell'}
= P_\ell \delta_{\ell \ell'}$.
These operators are defined in such a way that when operating
on any $f\in \calF$,
$P_\ell$ selects out those terms in the $q$-expansion of $f$ which
have powers equal to $\ell/4$ modulo unity -- \ie,
$$  (P_\ell f)\,[T]~=~ \exp \left\lbrace {{i\pi \ell}\over 2} \right\rbrace
    \, (P_\ell f) ~~~~~\forall\, f\in \calF~.
\eqn\projectionT
$$
For example,
if $f= \thetatwo^2 \thetathree^4 \thetafour^8$,
then
$$  \eqalign{
      P_1\,f ~&=~ \half \,\thetatwo^2 (\thetathree\thetafour)^4 \,
                  (\thetathree^4 ~+~ \thetafour^4)~, \cr
      P_3\,f ~&=~ \half \,\thetatwo^2 (\thetathree\thetafour)^4 \,
                  (\thetafour^4 ~-~ \thetathree^4)~, \cr
      P_0\,f ~&=~ P_2\,f ~=~ 0 ~.\cr}
\eqno\eq
$$
Let us also define (for the sake of typographical convenience)
the operator $\calS\equiv \exp\lbrace -4\pi i/K \rbrace [S]$.
Then our procedure is as follows.
Choose an $f\in \calF$ of the form $\vartheta^{26-D}$,
and calculate the quantity
$$  \bff ~\equiv~ \deltaminusone \,\lim_{n\to\infty}
    \left[ \half\,(1-P_1) (1+\calS)\right]^n
         \, f ~.
\eqn\bffeqn
$$
If $P_3\calS P_3 \bff =0$, then
up to one common scale factor
the corresponding solution for the $F$'s is:
$$
     F^1 ~=~ P_0 \calS P_3\,\bff~,~~~~~~
     F^2 ~=~ - P_2 \calS P_3\,\bff~,~~~~~~
     F^3 ~=~ P_3\,\bff~.
\eqn\solutionsF
$$

It is easy to see how this procedure works.
The goal is to construct a set of $F$'s which are distinguished
by their eigenvalues under $[T]$, and which furthermore are closed
under $[S]$.  This we achieve by constructing a quantity $\bff$ which
itself is $\calS$-invariant, and from which the individual $F$'s
can be obtained by projection.  The crucial element in $K=8$,
however, is the fact that this $\bff$
must satisfy $P_1 \bff=0$;  we cannot accommodate a fourth such $F$ with
$q$-powers equalling $1/4$ modulo unity in
building the desired partition functions [because there is no
corresponding term $D_8$ with this form in \ABCKequaleight].
We handle this difficulty as follows.
Starting from any $f$, a first ``guess''
for $\bff$ is the $\calS$-invariant quantity
$\bff^{(0)}=\half\deltaminusone (1+\calS)f$.  We then enforce our
requirement that $P_1 \bff=0$ by modifying the guess:  $\bff^{(1)}=
(1-P_1)\bff^{(0)}$.  However, we no longer are guaranteed that $\bff^{(1)}$
is $\calS$-invariant, and we therefore re-apply the operator $\half(1+\calS)$.
This process iterates until we have finally achieved an $\bff$ which
is $\calS$-invariant {\it and}\/ has no $P_1$-projection -- \ie,
which is preserved under applications of both operators $\half(1+\calS)$
and $(1-P_1)$.  We write this solution for $\bff$ formally as in \bffeqn\
above;  of course if this iterative process fails to converge one
must choose a new $f\in \calF$.  (In practice, however, with $f$
restricted to the space $\calF$ this process converges almost immediately.)
Having thus found $\bff$, we define the choice $F^3 \equiv P_3\bff$.
Since the $i=3$ component of \FconstraintS\ tells us that
$$  \calS \, F^3 ~=~ M_{(8)}^{13} \,F^1~+~ M_{(8)}^{23} \,F^2
                 ~=~  F^1 ~-~ F^2 ~,
\eqno\eq
$$
we must correspondingly define
$F^1= P_0 \calS F^3$ and $F^2=-P_2 \calS F^3$.
Note that these choices are consistent
only if $P_3 \calS F^3=0 $;  this follows from the fact that
$M_{(8)}^{33}=0$ in \Mmatrices.

Generating solutions for the $K=4$ case is even simpler.
Here we take the space $\calF$ of polynomials to be that
generated by $\lbrace \thetatwo^4,\thetathree,\thetafour\rbrace$
(so that $f[T^2]=f$ for all $f\in \calF$), and consequently we need
consider only the two projection operators $P_0$ and $P_2$ (since
$P_1 f=P_3 f=0$ for all $f\in \calF$).
For any chosen $f\in\calF$, we define
$$ \bff ~\equiv~ \half\,\deltaminusone\,(1+\calS)\,f ~,
\eqn\ffeqnfour
$$
whereupon we quickly have the solutions
(up to a common scale factor)
$$ F^1 ~=~ P_0\, \bff~,~~~~~~~~
   F^2 ~=~ 4\, P_2 \calS P_0 \,\bff~.
\eqn\Fsolutionsfour
$$

Note that there were two reasons this $K=4$ case was significantly
simpler than the $K=8$ case.  First, for $K=4$ we must
demand $P_1\bff=P_3\bff=0$
instead of the more difficult $K=8$ constraints
$P_1\bff=0$, $P_3\bff\not=0$;
the presence of {\it two} zero-constraints in $K=4$ instead of only one
allowed us to subsume them together into a restriction
in the space $\calF$.  Second, the matrix $M_{(4)}$ has no
zero entries;  hence the $F$'s can always be found by simple
projections and only their relative normalizations need be
determined.  In $K=8$, however, we must further assert $P_3\calS P_3\bff=0$;
as stated above,
this occurs because $M_{(8)}^{33}=0$.

Given these procedures for generating solutions for the $F$'s
satisfying \FconstraintS\ and \FconstraintT\ in
the $K=4$ and $K=8$ cases,
it is easy to build modular-invariant partition functions of
the proper forms.  Let us first construct some examples
for the $(1,4)$ case.  Taking $f=2 \beta^5$
[where we remind the reader of the definitions in \definitions],
we find
$\bff=\deltaminusone (\beta^5+\delta^5)$, whereupon
\Fsolutionsfour\ yields the results
$$  F^1~=~ \half \,\deltaminusone \,(\gamma^5+\delta^5)~,~~~~~~~~~
   F^2~=~\deltaminusone \,\lbrack 2\,\beta^5 ~+~ (\gamma^5-\delta^5)\rbrack ~.
\eqn\SOforty
$$
Since the internal gauge symmetry for such models is determined,
as usual, by the left-movers, we can quickly identify this solution
as corresponding to gauge group $SO(40)$, the largest allowed in
six spacetime dimensions.  As another example, we start with
$f=2 \beta^3(\beta^2+\gamma^2+\delta^2)$:  this yields
$\bff=\deltaminusone (\beta^3+\delta^3)(\beta^2+\gamma^2+\delta^2)$ and the
solutions
$$  F^1~=~ \half \,\deltaminusone
\,(\gamma^3+\delta^3)\,(\beta^2+\gamma^2+\delta^2)~,~~~~
    F^2~=~\deltaminusone \,\lbrack 2\,\beta^3 ~+~ (\gamma^3-\delta^3)\rbrack\,
     (\beta^2+\gamma^2+\delta^2) ~.
\eqn\SOtwentyfourEeigh
$$
The presence of the $E_8$ character $\beta^2+\gamma^2+\delta^2$ readily
identifies the gauge group corresponding to this solution
as $SO(24)\otimes E_8$.
As a third example for $K=4$, we can choose
$f=\gamma^2(\beta^3+\gamma^3+\delta^3)$.  From
this we obtain the solution
$$  \eqalign{
    F^1~&=~ \deltaminusone\, \left\lbrace
             \half\,(\gamma^3+\delta^3)\,(\gamma^2+\delta^2) ~+~ \half\,
               \beta^3\,(\gamma^2-\delta^2) \right\rbrace ~,\cr
    F^2~&=~ \deltaminusone\, \biggl\lbrace
        (\gamma^3+\delta^3)\,(\gamma^2-\delta^2) ~+~ \beta^3\,[2\beta^2
+(\gamma^2+\delta^2)] \cr
       &~~~~~~~~~~~~~~~+~ 2\,\beta^2\,(\gamma^3-\delta^3) \biggr\rbrace~,\cr}
\eqn\sotwofoursosixteen
$$
which (as we will see)
can be identified with the gauge group $SO(24)\otimes SO(16)$.
One can similarly construct partition functions for the $(1,8)$ theory:
for instance, starting with $f=2 \beta^{11/2}$, we
find (only one iteration required)
$\bff=\deltaminusone(\beta^{11/2}+\delta^{11/2})$,
which immediately leads to the solution
$$ \eqalign{
  F^1~&=~ \half \, \deltaminusone\, (\gamma^{11/2}~+~\delta^{11/2})~,~~~\cr
  F^2~&=~ \half \, \deltaminusone\, (\gamma^{11/2}~-~\delta^{11/2})~,~~~\cr
  F^3~&=~ \phantom{\half}\,\deltaminusone\, \beta^{11/2}~.\cr}
\eqn\sofortyfour
$$
Note that $P_3 \calS P_3\bff=0$, so this solution is indeed valid.
It is clear that this solution
corresponds to $SO(44)$, the largest allowed gauge group in $D=4$.

\section{Dictionaries for Model-Construction}

While thus far it has been quite straightforward to identify the gauge
groups corresponding to our partition functions, we have been
dealing only with the simplest of cases;  furthermore, in principle
almost any properly-chosen function $f$ can serve in generating
solutions, and we require a way to discern which of all possible
solutions correspond to bona-fide fractional
superstring {\it models}.
Toward this end we now develop a method for generating
partition functions which
we believe do precisely this in the heterotic $(1,K)$ cases,
and for which the underlying physics is substantially more transparent.
Our approach rests on two fundamental observations.

The first observation has to do with the existence of a model
with maximal gauge symmetry for the $(1,2)$ heterotic string
in $D$ dimensions.  For any value $D\leq 10$,
there is always a self-consistent $(1,2)$ model which can be
formulated with gauge group $SO(52-2D)$ -- such a model is,
in a sense, the starting point in model-building, for all other
models can be obtained from it by altering this known solution
via orbifolding, \eg, by adding twists, altering the boundary
conditions of worldsheet fields,
adding new sectors, \etc, all of which tends to
break the gauge group and correspondingly add new terms
to the partition function.
We therefore assert that such a self-consistent
model exists as well for the $(1,K)$ fractional superstring,
and has gauge group $SO(52-2D)= SO(48-16/K)$;
indeed, the validity of this assertion follows directly from the
(assumed) self-consistency of the $K>2$ fractional superstring
right-moving sector
and the near-decoupling of the gauge sector (the $K=1$ left-moving
sector), as will be discussed below.
Since we have already constructed the unique partition functions
corresponding to these gauge groups [see \SOforty\ and \sofortyfour],
it therefore follows that these two partition
functions must indeed correspond to actual $(1,K)$ models.

Our second and more important observation concerns the couplings between
the left- and right-moving sectors of a heterotic-type string theory.
As is well-known for the traditional $(1,2)$ heterotic string
theory, the left-moving sector carries with it all of the information
concerning the internal gauge group and particle representations;  the
right-moving sector, on the other hand,
carries with it the linkage to spacetime
physics, Lorentz spin and statistics, and spacetime supersymmetry.
One builds a model, then, by choosing these respective sectors so
that certain physical constraints are satisfied:
one must maintain worldsheet (super)conformal
invariance, modular invariance (which incorporates proper level-matching),
spacetime Lorentz invariance, and
physically sensible internal (GSO-like) projections (thereby incorporating
proper spacetime spin-statistics).
Of course, not all of these requirements are independent.
Some of these requirements can clearly be
placed on the left- and right-moving sectors separately:
among these are, for example, worldsheet (super)conformal invariance,
spacetime Lorentz invariance, and physically sensible projections.
Modular invariance, on the other hand, constrains both sectors jointly.
Thus, when building a model, one must satisfy essentially two
kinds of constraints:  those which involve the left- and right-moving
sectors of the theory independently (guaranteeing that they
are each internally self-consistent), and those
(\ie, modular invariance and the implied level-matching)
which insure that they are properly coupled or linked
to each other.

In the case of the heterotic $(1,K)$ fractional superstring,
we expect the same situation to prevail:
we must determine those $K=1$ left-moving sectors which
are themselves internally self-consistent, and then we must join them
with our (assumed self-consistent) $K>2$ right-moving sector
in such a way that modular-invariance is satisfied.
Fortunately, the underlying physics of a $K=1$ left-moving
sector is well-understood;  for example, descriptions of it
in terms of lattices, orbifolds, or Fock-space spectrum-generating
formulae abound.
In particular, it is well-known how to construct valid $(1,2)$
models which satisfy {\it all}\/ of the physical
model-building constraints we have listed.  Therefore, one
might hope to be able to build valid $(1,K)$ models by
first building valid $(1,2)$ models, and then
``replacing'' their $K=2$ right-moving sectors with
our new $K>2$ right-moving sectors in such a way that
modular-invariance (the sole ``linking''
constraint) is not violated.  Such a procedure would
thereby guarantee, in the language of the previous subsection,
a set of $F^i$'s which themselves are known to correspond to
valid $K=1$ left-moving sectors.

It turns out that
these arguments can be phrased directly in terms of a correspondence
or ``dictionary''
between right-moving $K=2$ physics and $K>2$ physics,
in the sense that they may be substituted for each other
in this way when building models.
At the level of the partition function (which has been the
basis of our approach), this means that we
are able to draw a correspondence between the respective
$\Theta$-functions of these right-moving sectors.
For the $K>2$ theory
these $\Theta$-functions are simply the expressions $A^i_{(K)}$
which we have been using to build our partition functions, and for the $K=2$
case these $\Theta$-functions are the usual
Jacobi $\vartheta$-functions.  We therefore
expect to be able to construct a dictionary relating the expressions
$A^i_{(K)}$ to the
Jacobi $\vartheta$-functions, in the sense that the two
underlying sectors giving rise to these respective expressions
couple in the same manner to left-moving physics,
and hence can be used interchangeably
in building self-consistent models of either the $(1,2)$ or
$(1,K)$ type.

It is fairly straightforward to construct this dictionary
between the $A^i_{(K)}$ functions and the $\vartheta$-functions,
for our first assumption [the validity of the $(1,K)$ solutions
\SOforty\ and \sofortyfour]
allows us to make this connection in the case of maximal gauge
symmetry.
Let us first concentrate on the case $K=4$.  Recall from Sect.~III
that the supersymmetric $(1,2)$ model in $D=6$ with the maximal gauge
group $SO(40)$ has the partition function
$$  \calZ_{SO(40)} ~=~ (\half)^2\,\tautwo^{-2} \,\calK\, (
    \bbar \beta^5+\cbar \gamma^5+\dbar \delta^5)
\eqn\recallsoforty
$$
(where $\calK\equiv \deltaminusone \deltabarminushalf \Jbar$).
Repeatedly making use of the algebraic identity
$$  Aa +Bb ~=~ \half(A+B)(a+b) ~+~
                        \half(A-B)(a-b) ~,
\eqno\eq
$$
it turns out that we can write
$$ \eqalign{
   \bbar \beta^5 +\cbar \gamma^5 +\dbar \delta^5~&=~
      \phantom{+~}\quarter \, (\bbar+\cbar-\dbar)
   \,[2\beta^5+(\gamma^5-\delta^5)] \cr
     &\phantom{=~}+~ \quarter \, (\bbar-\cbar+\dbar)
    \,[2\beta^5-(\gamma^5-\delta^5)] \cr
     &\phantom{=~}+~ \half    \, (\cbar+\dbar) (\gamma^5+\delta^5) ~.\cr }
\eqn\intermediatestep
$$
We therefore have
$$ \eqalign{
   \calZ_{SO(40)} ~&=~ \quarter\,\tautwo^{-2} \,\deltaminusone
      \deltabarminushalf\,\Jbar \, \biggl\lbrace
     \quarter \, (\bbar+\cbar-\dbar) \,[2\beta^5+(\gamma^5-\delta^5)] \cr
     &~~~~~~~~~~~+~\half    \, (\cbar+\dbar) (\gamma^5+\delta^5)
     ~-~\quarter \, \Jbar \,[2\beta^5-(\gamma^5-\delta^5)] \biggr\rbrace\cr}
\eqn\Zrewritten
$$
for the $(1,2)$ maximally symmetric model in $D=6$.

The next step involves properly interpreting some
of the factors in \Zrewritten.
Since this is the partition
function of a $(1,2)$ model, the critical dimension is 10, and reducing
the spacetime dimension to 6 through fermionization
(as discussed in Sect.~III) yields worldsheet matter
consisting of the 4 transverse bosonic coordinate fields,
16 Majorana-Weyl right-moving fermions, and 40
Majorana-Weyl left-moving fermions.
This is why each term in \Zrewritten\ contains
two anti-holomorphic powers of $\bbar$, $\cbar$, or $\dbar$
(recall that $J\equiv \gamma-\beta-\delta$), and five holomorphic powers
of $\beta$, $\gamma$, or $\delta$.
Of these 16 right-moving fermions,
four are the superpartners of the bosonic coordinate fields
(and hence carry spacetime Lorentz indices)
while the remaining twelve are internal and carry only
internal quantum numbers.
However, of these twelve, four
had previously been the superpartners of the (now fermionized)
$10-6$ coordinate bosons, and as such their degrees of
freedom (in particular, their toroidal boundary conditions)
must be chosen to be the same as those of
the four fermions carrying Lorentz indices.
In Ref.~[\KLT], for example,
this is the result of the so-called ``triplet'' constraint,
which arises by demanding
the periodicity or anti-periodicity of the worldsheet supercurrent
on the worldsheet torus.
Thus, the 16 right-moving Majorana-Weyl fermions split into
two groups:  the first eight (which include the four with
spacetime Lorentz indices) must all have the same boundary conditions,
and the remaining eight (all of which are are internal fermions)
can be chosen independently.
As we can see from the partition functions above,
the first group of eight right-moving
worldsheet fermions are
precisely the ones which combine to produce the overall factor
of $\Jbar$:  this is indeed the expression of spacetime supersymmetry.
The remaining eight right-moving fermions, however, are
all internal, and these are responsible for the
remaining single powers of $\bbar$, $\cbar$, and $\dbar$
within the braces in \Zrewritten\ above.

This understanding is very important, for it enables
us to interpret the $\Jbar$ factor in the second line of $\Zrewritten$.
As we stated, the $\Jbar$ in the first line expresses
spacetime supersymmetry:  the identity $\Jbar=0$ represents
the complete cancellation of spacetime bosonic states
against spacetime fermionic states.
The {\it second}\/ factor of $\Jbar$, however, arises
from exclusively internal degrees of freedom, and thus for
this term the identity $\Jbar=0$ represents an
internal GSO-like projection between particles of the {\it same}
spacetime statistics (in fact, between the same particle states).
Thus, the last term in \Zrewritten\ contains no physical states whatsoever,
and may be legitimately dropped.
We therefore obtain
$$ \eqalign{
   \calZ_{SO(40)} ~&=~ \quarter\,\tautwo^{-2} \,\deltaminusone
      \deltabarminushalf\,\Jbar \,\times \cr
   &~~~\times\, \biggl\lbrace
     \quarter \, (\bbar+\cbar-\dbar) \,[2\beta^5+(\gamma^5-\delta^5)] ~+~
     \half    \, (\cbar+\dbar) (\gamma^5+\delta^5) \biggr\rbrace\cr}
\eqn\ZrewrittenJgonepre
$$
for this $(1,2)$ model, or equivalently
$$  \calZ_{SO(40)} ~=~ \tautwo^{-2}\,
    \biggl\lbrace \overline{R_1^{K=2,D=6}} \,  L_1^{SO(40)}
  ~+~ \overline{R_2^{K=2,D=6}} \, L_2^{SO(40)} \biggr\rbrace
\eqn\ZrewrittenJgone
$$
where the two factors arising from the right-moving $K=2$, $D=6$ theory
are
$$     \eqalign{
   R_1^{K=2,D=6} ~&\equiv~ \,{\textstyle{1\over 4}}\,
     \deltaminushalf \,J\,(\gamma+\delta)~\cr
   R_2^{K=2,D=6} ~&\equiv~ {\textstyle{1\over {16}}}\,
    \deltaminushalf \,J\, (\beta+\gamma-\delta)~,\cr }
\eqn\Rfactorsfour
$$
and where the two factors arising from the left-moving $K=1$ $SO(40)$
theory are:
$$  \eqalign{
   L_1^{SO(40)}~&\equiv~ \half\,\deltaminusone\,(\gamma^5+\delta^5)\cr
   L_2^{SO(40)}~&\equiv~\phantom{\half}\,
    \deltaminusone\,(2\beta^5+\gamma^5-\delta^5) ~.\cr}
\eqn\Ffactorsfour
$$

Now that we have rewritten \recallsoforty\ in the form
\ZrewrittenJgone, we can compare this result with
the partition function for the
$(1,4)$ $SO(40)$ model.
Recall that this latter partition function was found
to be [see \fs\ and \SOforty]:
$$  \eqalign{
      \calZ ~&=~ \tautwo^{-2}\,\biggl\lbrace
        \half\,\overline{A_4}\,\deltaminusone\,(\gamma^5+\delta^5)
   ~+~ \overline{B_4}\,\deltaminusone
      \,[2\beta^5+(\gamma^5-\delta^5)]\biggr\rbrace~\cr
  &=~\tautwo^{-2}\,\biggl\lbrace  \overline{A_4}\,L_1^{SO(40)} ~+~
       \overline{B_4}\,L_2^{SO(40)}\biggr\rbrace~;\cr }
\eqn\recallit
$$
note that one can verify the overall normalization of \recallit\
by counting the numbers of low-lying (\eg, tachyonic
or massless fermionic) degrees of freedom.
It is clear that \ZrewrittenJgone\ and \recallit\ have the same
left-moving holomorphic pieces $L_i^{SO(40)}$, and therefore we can
relate them to each other, \ie,
$$
 \calZ_{(1,2)}~\equalstar~\calZ_{(1,4)} ~~~~~~~ {\rm for~} SO(40)
\eqn\identifyPF
$$
(where $\equalstar$ indicates this relation or correspondence),
if we make the following correspondences:
$$  \eqalign{
     A_{(4)}^1~\equiv~A_4~&\equalstar~
       R_1^{K=2,D=6} \cr
     A_{(4)}^2~\equiv~B_4~&\equalstar~
       R_2^{K=2,D=6} \cr}
\eqn\dictionaryfour
$$
where the $R$'s are given in \Rfactorsfour.
This result in \dictionaryfour, then,
is our $K=4$ ``dictionary'' relating $A_4$ and $B_4$
to the Jacobi $\vartheta$-functions, thereby enabling
us to interchange their corresponding $K=2$ and $K=4$ right-moving sectors
(at the partition-function level) in order to build models of the $(1,4)$
variety.
We remark that in principle it is just as straightforward
to construct dictionaries suitable for building $(2,4)$ models;
indeed, the expressions $R_i^{K=2,D=6}$ which we have obtained can
themselves be taken as the partition-function contributions
from the self-consistent {\it left-moving} sectors of such theories.

Note that if we had {\it not}\/ dropped the
third term in the $(1,2)$ partition function \Zrewritten,
we would have required a {\it third} string-function expression $C_4$
to relate to
$$     R_3^{K=2,D=6}~\equiv~\deltaminushalf\, J \,(\gamma-\beta-\delta )~=~
     \deltaminushalf\, J_{\rm st}\, J_{\rm int}~.
\eqn\badterm
$$
Here we have explicitly indicated the
origins of the two independent Jacobi factors, labelling with
subscripts whether they arose in the $(1,2)$ $D=6$ theory
from spacetime or internal degrees of freedom.
However, such an additional
string-function expression $C_4$ does not exist.
It is indeed fortunate that the GSO-projection $J_{\rm int}=0$
enables us to avoid this unwanted term $R_3$ in an internally self-consistent
manner, and thereby obtain the dictionary \dictionaryfour.

There are several things to note about this dictionary.
First, we observe that this dictionary is self-consistent
as a relation between modular functions,
with each side of the relation transforming under the modular
group with weight $k= -2$ [as appropriate for $D=6$, according
to \kDformula].
Furthermore, the expressions on the right side of
\dictionaryfour\ mix under the $[S]$-transformation
with the same matrix $M_{(4)}$ found for the
$A^i_{(4)}$, and the same, of course,
holds true for the $[T]$-transformation and the $N_{(4)}$
matrix.\foot{
    One must not forget that when performing the $[S]$
    transformation on the right sides of \dictionaryfour,
    self-consistency requires setting factors of $J_{\rm int}$ to zero.}
Indeed, the right sides of \dictionaryfour\ each
have $q$-expansions of the forms
$q^h(1+...)$ where inside the parentheses all $q$-powers
are integral and where $h$-values are equal (mod 1)
on both sides of each equation.
Additionally, this dictionary incorporates the spacetime supersymmetry
of $A_4$ and $B_4$ in a natural way, allowing these expressions
(which are themselves the ``Jacobi identities'' for $K=4$)
to correspond to expressions proportional to the spacetime factor $J$,
the $K=2$ Jacobi identity.

In fact, from this dictionary it is now clear that one cannot
expect to factorize the string-function expressions $A^i_{(4)}$
into two pieces, one of which might vanish on its own.
In the $(1,2)$ theory [\ie, the right side of \dictionaryfour],
the $J$ factor is the result of spacetime-related degrees of freedom,
and the remaining factors linear in $\beta$, $\gamma$, and $\delta$
in \dictionaryfour\
are the contributions from the internal right-moving degrees of
freedom.  For the $K=4$ theory, however, we are {\it in}
the critical dimension, and hence {\it all} right-moving modes
carry spacetime information and play a part in yielding spacetime
supersymmetry.
We therefore do not expect to be able to factorize the left sides
of \dictionaryfour\ in order to achieve separate correspondences
between individual factors on both sides of the equations.

This dictionary allows us to easily generate $(1,4)$ fractional
superstring models:  we start with a known $(1,2)$ model compactified
in some manner to six spacetime dimensions, and then make the
``translation'' given in \dictionaryfour\ to
the $K=4$ right-moving sector.  As examples, we can translate the remaining
six-dimensional $(1,2)$ models whose partition functions were
given in \Dequalsixmodels.
The $SO(24)\otimes E_8$ model is particularly simple to translate,
because in the present case the analogue of \intermediatestep\
becomes
$$ \bbar \beta^3 +\cbar \gamma^3 +\dbar \delta^3~=~
     \quarter \, (\bbar+\cbar-\dbar) \,[2\beta^3+(\gamma^3-\delta^3)]~
     +~ \half    \, (\cbar+\dbar) (\gamma^3+\delta^3) ~
\eqn\intermediatesteptwo
$$
where we have set $\Jbar_{\rm internal}=0$.
Upon substituting this expression
into the original partition function in \Dequalsixmodels\ and
translating according to \dictionaryfour, we indeed find
the partition function \fs\ with the $F$'s given in \SOforty.
This therefore confirms that the solution in \SOforty\
corresponds to a valid $(1,4)$ model, and has the (left-moving)
gauge group claimed.
Similarly translating the $SO(24)\otimes SO(16)$ partition
function in \Dequalsixmodels,
we obtain the solution quoted in \sotwofoursosixteen, again
confirming the interpretation of that solution as corresponding
to a valid model with the quoted gauge group.

We stress again, of course, that this dictionary derivation
does not merely duplicate the results found earlier, for
the general procedure presented in the previous subsection
merely assures the creation of modular-invariant expressions $\calZ$.
It is the crucial fact that we can
derive these particular solutions {\it via our dictionary translation}
which guarantees their interpretation as the partition functions
of actual models.
To illustrate this fact, let us consider the creation
of $(1,4)$ models having gauge groups $G$ of the form
$G_{12}\otimes G_8$, where
$G_r$ denotes a simple (and simply-laced) group of rank $r$.
Recall that the procedure given in the previous subsection allows us
to specify our possible $(1,4)$ functions $\calZ$ in terms of the simpler
$\calS$-invariant expression $\bff$.
Since we expect a gauge group
$G_r$ to reveal itself in $\calZ$
through the presence of factors which are linear combinations of
$\beta^p$, $\gamma^p$, and $\delta^p$ where $p=r/4$ [for example, the
factor (or character) corresponding to $E_8$ is $(\beta^2+\gamma^2+\delta^2)$],
we can survey many such $\calZ$'s by building functions $\bff$
of the form $\bff^{(i,j)}=\deltaminusone Q_3^{(i)} Q_2^{(j)}$ where,
for example,
$$  Q_p^{(1)}= \gamma^p~, ~~~~Q_p^{(2)}= \beta^p+\delta^p~,~~~~
    Q_p^{(3)}= \beta^p+\gamma^p+\delta^p~, ~~~~Q_p^{(4)}=
\beta^p-\gamma^p+\delta^p~.
\eqn\Qdefs
$$
Some of these possibilities we have already examined:
for example, we have confirmed that $\bff^{(3,1)}$
generates an $SO(24)\otimes SO(16)$ model, and
that $\bff^{(1,3)}$, $\bff^{(2,3)}$, and $\bff^{(3,3)}$
each generate the $SO(24)\otimes E_8$ model.
In fact, we have also seen that $\bff^{(1,1)}$
leads to the maximally-symmetric $SO(40)$ model (by construction),
and $\bff^{(4,3)}$ turns out to be a null solution (all the $F$'s vanish).
However, there are eight other distinct modular-invariant
functions $\calZ$ which can be constructed in this form
[eight rather than ten because $\bff^{(4,1)}$, $\bff^{(4,2)}$,
and $\bff^{(4,4)}$ each lead to the same $\calZ$],
and each of these might reasonably correspond to a valid
$(1,4)$ model.  For instance, $\bff^{(4,1)}$
leads to a modular-invariant function $\calZ$
which is the translated version of
$$   \calZ ~=~ \tautwo^{-2}\, \calK\,
       (\bbar \beta^2-\cbar \gamma^2+\dbar \delta^2)\,
               (\beta^3-\gamma^3+\delta^3)~.
\eqn\nomodel
$$
However, there does not exist a $(1,2)$ model having \nomodel\
as its partition function.
Our dictionary therefore enables us to conclude that
there is no $(1,4)$-type model which can be generated for
this case.

In a similar manner, we can obtain the corresponding dictionary
for $K=8$ ($D=4$).  In four spacetime dimensions the maximal
left-moving gauge group is $SO(44)$, and the partition function of
the $(1,2)$ $SO(44)$ model was given in \fourdimsofourfour.
It is a simple matter to rewrite
$$ \eqalign{
    \bbar^{3/2} \beta^{11/2} &~+~
    \cbar^{3/2} \gamma^{11/2} ~+~
    \dbar^{3/2} \delta^{11/2} ~=~ \cr
    &=~\bbar^{3/2} \beta^{11/2} ~+~
     \half\,(\cbar^{3/2}+\dbar^{3/2})\,(\gamma^{11/2} +\delta^{11/2}) ~+~ \cr
     &~~~~~+~ \half\,(\cbar^{3/2}-\dbar^{3/2})\,(\gamma^{11/2}
-\delta^{11/2})~, \cr}
\eqno\eq
$$
whereupon a quick comparison with the solution found in \sofortyfour\
yields the $K=8$ dictionary:
$$  \eqalign{
     A_{(8)}^1~\equiv~A_8~&\equalstar~
      R_1^{K=2,D=4} \cr
     A_{(8)}^2~\equiv~B_8~&\equalstar~
      R_2^{K=2,D=4} \cr
     A_{(8)}^3~\equiv~C_8~&\equalstar~
      R_3^{K=2,D=4} \cr }
\eqn\dictionaryeight
$$
where the factors from the right-moving $K=2$ $D=4$ theory are:
$$  \eqalign{
     R_1^{K=2,D=4}~&\equiv~
      \quarter\,\deltaminushalf \, J\, (\gamma^{3/2}+\delta^{3/2}) \cr
     R_2^{K=2,D=4}~&\equiv~
      \quarter\,\deltaminushalf \, J\, (\gamma^{3/2}-\delta^{3/2}) \cr
     R_3^{K=2,D=4}~&\equiv~
      \quarter\,\deltaminushalf \, J\, \phantom{(}\beta^{3/2} ~.\cr }
\eqn\Rfactorseight
$$
Note that it is extremely fortunate that the solution
\sofortyfour\ could be constructed satisfying $P_3SP_3\bff=0$, for
if this solution had not existed, the $K=8$ dictionary could
not have been built.
Once again, we observe that this dictionary is self-consistent,
with both
sides of \dictionaryeight\ having modular weight
$k= -1$ and mixing identically according the matrices $M_{(8)}$
and $N_{(8)}$ under the $[S]$- and $[T]$-transformations respectively.
We remark that $(2,8)$ models can be built
as well, simply by taking these expressions $R_i^{K=2,D=4}$
as the valid representative partition-function contributions from
self-consistent {\it left-moving} $K=2$ sectors.

As in the $K=4$ case, the $J$ factors in \Rfactorseight\
arise in the $(1,2)$ theory from spacetime-related degrees of freedom;
the remaining factors, on the other hand, arise from purely internal
degrees of freedom.  However, unlike the $K=4$ case,
it would have been impossible now to obtain a complete GSO projection
with these internal factors, for the analogous combination
$\gamma^{3/2}-\beta^{3/2}-\delta^{3/2}$ is non-vanishing.
Fortunately, the $K=8$ theory provides {\it three} string-function
expressions $A_8$, $B_8$, and $C_8$ to relate to our three $R^{K=2,D=4}$
factors,
so our $K=8$ dictionary could nevertheless be constructed.
It is indeed curious and fortuitous that the
$K=4$ and $K=8$ theories provide exactly the
needed number of independent string-function expressions
with which to build supersymmetric fractional superstring
partition functions.

\endpage

\chapter{MODEL TRANSLATABILITY}

In the previous section we established our dictionaries
and discussed how they
enable us to confirm whether the possible functions $\calZ_{(1,K)}$
correspond to valid $(1,K)$ models.
Of considerably more interest, however, is the question
of examining those that {\it do} correspond to models,
for we would like to examine this space of new $(1,K)$ models
and determine, for example, its size and other relevant features.
This is, in a sense, the opposite issue, for now
we must use our dictionaries to determine which of the
$(1,2)$ models may indeed be translated.
We will find that not all $(1,2)$ models may be translated,
and that only those possessing a {\it maximal}\/ number of spacetime
supersymmetries are in correspondence with valid $(1,K)$ models
(which themselves have $N=1$ spacetime supersymmetry\refmark{\letter}).

At first glance it may seem that our dictionaries for $K=4$ and $K=8$
allow {\it any} spacetime-supersymmetric $(1,2)$
model to be translated.  Indeed, for the $K=4$ case,
if our $(1,2)$ model has a partition function $\calZ$ of the form
$$   \calZ_{(1,2)} ~=~
        \tautwo^{-2}\, \deltaminusone\,\deltabarminushalf\,\Jbar\,
        (\bbar X + \cbar Y + \dbar Z)
\eqn\translatableform
$$
where $X$, $Y$, and $Z$ are arbitrary holomorphic $\vartheta$-function
expressions, then the algebraic identity
$$  \eqalign{
        \bbar X + \cbar Y +& \dbar Z ~=~
   (\cbar+\dbar)\,\left[\half(Y+Z)\right] \cr
 &+~(\bbar+\cbar-\dbar) \,\left[\half X+\quarter Y-\quarter Z\right] ~-~
      \Jbar\,\left[\half X - \quarter Y + \quarter Z\right] \cr}
\eqn\XYZidentity
$$
and the dictionary \dictionaryfour\ allow us to construct the
partition function of the corresponding $(1,4)$ model:
$$  \calZ_{(1,4)} ~=~ \tautwo^{-2}\, \deltaminusone\,\biggl\lbrace
       \overline{A_4}\,(Y+Z) ~+~
       \overline{B_4}\,[4\,X + 2\,(Y-Z)] \biggr\rbrace ~.
\eqn\genlsolution
$$
This result will be modular-invariant provided \translatableform\
is modular-invariant.
Note that the identity \XYZidentity, which makes the
necessary rewriting possible, is simply the statement
that the linear combinations $\cbar+\dbar$, $\bbar+\cbar-\dbar$, and
$\Jbar\equiv \cbar-\bbar-\dbar$
span the three-dimensional space spanned by the combinations
$\bbar$, $\cbar$, and $\dbar$ separately, and therefore setting
$\Jbar=0$ (the internal GSO projection)
always leaves the remaining two left-moving factors $F^1$ and $F^2$ to
multiply the right-moving factors
$\overline{A_4}$ and $\overline{B_4}$ in the $(1,4)$ partition function.
The same argument can be made for the $K=8$ case as well.
It would therefore seem that the spaces of $(1,K)$ models
are as large as the spaces of $(1,2)$ models compactified
to the appropriate dimensions $D=2+16/K$, a conclusion
which would suggest that the dictionary translation idea
does not yield the expected substantial truncation in the
sizes of the space of $(1,K)$ models (which are of course {\it in}
their critical dimensions).

Fortunately, this is not the case, for there are two important
reasons why valid supersymmetric $(1,2)$ models may fail to be translatable.
First, they may fail to have partition functions of the needed
general forms
$$   \calZ ~=~ \tautwo^{-2}\, \deltaminusone\,\deltabarminushalf\,\Jbar\,
       \biggl\lbrace\bbar^{(10-D)/4} X + \cbar^{(10-D)/4} Y +
    \dbar^{(10-D)/4} Z\biggr\rbrace ~;
\eqn\translatableformgenl
$$
indeed, we will see that the vast majority
of known supersymmetric $(1,2)$ models do not have this form.
Second, even though a given model may technically have a partition
function of the form \translatableformgenl, its separate
bosonic and fermionic contributions to that partition function
may themselves fail to be of the correct forms.  We will discuss each
of these possibilities in turn.

The first possibility is that a given $(1,2)$ spacetime-supersymmetric
model may fail to have a partition function of the form
\translatableformgenl.  While the assumed spacetime supersymmetry guarantees
the presence of the factor $\Jbar$, and the $\Delta$-functions
must always appear as in \translatableformgenl\ for a heterotic
$(1,2)$ string theory, we are not assured
that our remaining right-moving (anti-holomorphic) factors
will be $\bbar^{(10-D)/4}$, $\cbar^{(10-D)/4}$, $\dbar^{(10-D)/4}$,
or their linear combinations.
Indeed, from \onetwotheoryarbdim\ and \powers\ we see that
we are assured only that these remaining factors must contain $10-D$
powers of $\overline{\vartheta}$-functions;  we could thus in principle obtain
(in the $K=4$ case) expressions such as $(\bbar\cbar)^{1/2}$,
$(\bbar\dbar)^{1/2}$,
and $(\cbar\dbar)^{1/2}$ appearing instead.
Since the self-consistency of the underlying
$(1,2)$ model demands that the number of remaining $\vartheta$ powers
be even, we see that the above fractional powers are indeed
our only other possibilities for $D=6$, but most supersymmetric $(1,2)$
models in $D=6$ will involve these other factors in their partition functions.
The same problem exists for the $D=4$ case as well:
here translatability requires that
these factors take the forms $\bbar^{3/2}$,
$\cbar^{3/2}$, or $\dbar^{3/2}$,
yet terms such as $\bbar\cbar^{1/2}$ and $(\bbar\cbar\dbar)^{1/2}$
are also legitimate (and in fact common) occurrences in
the space of supersymmetric models.
We therefore must understand which models do not have such
factors in their partition functions, and will thereby be translatable.

It is clear that in order to obtain the desired remaining
anti-holomorphic factors --and simultaneously avoid the
unwanted ones-- our underlying model must have a symmetry
relating the internal (\ie, non-spacetime-related) right-moving
worldsheet fermions so that they each uniformly produce the same
contributions to $\calZ$.
For example, these $10-D$ internal fermions must have the same toroidal
boundary conditions in all sectors of the model, and be
in all respects interchangeable.
Such a worldsheet symmetry is not new:
if such a symmetry appears amongst the {\it left-moving}
worldsheet fermions, there will be corresponding massless spacetime
vector particles transforming in the adjoint representation of this symmetry
group.  Thus, such a left-moving worldsheet symmetry
can be interpreted as a spacetime gauge symmetry.
What, however, are the spacetime consequences of such
a {\it right-moving} worldsheet symmetry involving these
$10-D$ worldsheet fermions?

Fortunately, such a worldsheet symmetry is easy to interpret:
it is responsible for a multiplicity in the number of spacetime
gravitinos, so that the larger the rank of the symmetry group,
the larger the multiplicity.
In fact, if the rank of this worldsheet symmetry group is $10-D$ (so
that {\it all}\/ internal right-moving fermions are involved,
as is needed for translatability), then the
number of gravitinos in the spectrum of the model is $\Nmax$, where
$\Nmax$ is given in \nmax.
The analysis needed for proving this assertion
is not difficult, but varies greatly with the type
of $(1,2)$ model-construction procedure we employ;
in Appendix~B we provide a proof using the
free-fermion construction of Ref.~[\KLT].
We therefore conclude that only models with $N=\Nmax$ supersymmetry
have partition functions of the form \translatableformgenl.\foot{
   This statement assumes that we have carefully
   avoided cancelling spacetime bosonic and fermionic states in constructing
   the partition function, as will be discussed below.}

In the $K=2$ case ($D=10$), this of course produces a trivial
result:  since $\Nmax=1$ for $D=10$, the only models
which have partition functions of the form \translatableformgenl\
are indeed those which are supersymmetric.
There are precisely two such models,\refmark{\heteroticrefs}
and their partition functions are given in \Dequaltenmodels.
In the $K>2$ cases, however, this translatability requirement
is much more drastic, for the great majority of supersymmetric
$(1,2)$ models in $D<10$ do not have $N=\Nmax$ SUSY.
For example, it has been
found\REF\DS{D.~S\'en\'echal, {\it Phys.~Rev.} {\bf D39},
3717 (1989);  D.~S\'en\'echal, Ph.D.~thesis, Cornell University
(August 1990).}\refend\
that among a certain broad class of
$D=4$ free-fermion models, there exist fewer than 1150 with $N=4$ SUSY;
this is to be compared against over $32\,000$ models with $N=1$
SUSY\refmark{\DS}\
and a virtually limitless supply with no spacetime supersymmetry at
all.\REF\KRD{See, \eg,
K.R.~Dienes, {\it Phys.~Rev.~Lett.} {\bf 65}, 1979 (1990),
where more than $120\,000$ of such models have been explicitly
constructed.}\refend\
We see, therefore, that the number of translatable $(1,2)$
models in $D=4$ ---and hence the number of $(1,8)$ models---
is severely restricted and certainly under $1500$.
A similar restriction exists for the $K=4$ case as well.

As mentioned earlier,
there is also a second reason why a $(1,2)$ model may fail
to be translatable, even if it does technically have a partition
function of the algebraic form \translatableformgenl.
Let us consider the separate contributions
to the total partition function of a model
from the {\it spacetime} bosonic and fermionic states respectively.
Recall that in the spacetime Jacobi factor $\Jbar$, the term $(\cbar-\dbar)$
is the contribution from spacetime bosons (the Neveu-Schwarz sector)
and the term $\bbar$ is the contribution from spacetime fermions
(the Ramond sector);  the other anti-holomorphic factors in the partition
function represent purely internal degrees of freedom.
Let us therefore separate these two pieces, and write the
general bosonic and fermionic partition functions from an arbitrary
model as follows:
$$ \eqalign{
  \calZ_{b}~&=~ \tautwo^k\,\deltaminusone\,\deltabarminushalf\,
      \biggl\lbrace (\cbar -\dbar) \,V
        ~+~ W_{b} \biggr\rbrace~,\cr
  \calZ_{f}~&=~ \tautwo^k\,\deltaminusone\,\deltabarminushalf\,
      \biggl\lbrace \bbar \,V ~+~ W_{f}
      \biggr\rbrace~.\cr}
\eqn\bosonfermionpieces
$$
Here $W_b$ is that part of $\calZ_b$ whose spacetime anti-holomorphic
factor is not proportional to $(\cbar-\dbar)$.
It is clear that the total partition function
$\calZ\equiv\calZ_{b} -\calZ_f$
will be of the algebraic form \translatableformgenl\
if $W_b=W_f\equiv W$
where $W$ is arbitrary and if $V$ is of the form of the term in braces
in \translatableformgenl.
However, translatability additionally requires that $W$ vanish, regardless of
its form.  The reason for this is quite simple.  If $W\not=0$, then
allowing $W_b$ and $W_f$ to cancel in the difference $\calZ_b-\calZ_f$
amounts to cancelling spacetime bosonic and fermionic states
and is therefore inconsistent with our retention of the spacetime Jacobi factor
$\Jbar$ in \translatableformgenl;   indeed, writing $W_b-W_f=0$
is tantamount to simply writing $\Jbar=0$ in \translatableformgenl.
Another way of seeing this is to realize that while $W_b$ and $W_f$
might be algebraically equal, the {\it spacetime} factor in $W_b$
must include combinations of $\cbar$ and $\dbar$ only (since it comes
from spacetime bosonic degrees of freedom), while
the spacetime factor in $W_f$ must be simply $\bbar$.  Such expressions
$W_b$ and $W_f$ should therefore not be cancelled, so translatability
requires that they not appear at all.
Note that this additional constraint $W=0$ is not vacuous, for there exist
many models for which the total $\calZ$
is of the algebraic form \translatableformgenl\ but for which
$W\not=0$:
these are models with partition functions of the form $\calZ+W-W$.
It turns out, however, that any $(1,2)$ model with $N=\Nmax$ SUSY
is guaranteed to have $W=0$, so the $N=\Nmax$ constraint
remains sufficient to ensure model translatability.  (In fact, as
noted earlier, only by requiring $W=0$ does the $N=\Nmax$ constraint
cease to be {\it overly} restrictive.)
Thus, we see once again that $N=\Nmax$ SUSY is the $(1,2)$ physics
underlying translatability.

One might argue that this $N=\Nmax$ translatability constraint
is somewhat artificial:  our dictionary was constructed by
comparing the $SO(48-16/K)$ $(1,K)$ model with the maximally
symmetric $SO(52-2D)$ $(1,2)$ model, and since this latter
model always has an $N=\Nmax$ right-moving sector, this
$N=\Nmax$ constraint was thereby ``encoded'' into our dictionary
from the beginning.  Indeed, one might claim that other
dictionaries could be constructed for $N<\Nmax$ cases
simply by comparing, for example, our maximally symmetric
$(1,K)$ model with a maximally symmetric $N<\Nmax$ $(1,2)$ model.
However, such approaches ultimately fail to yield self-consistent
dictionaries.  At a mathematical level, this occurs
because it is not possible to construct
alternate $\vartheta$-function expressions which
could appear on the $K=1$ sides of such dictionaries:
such expressions would have to be not only eigenfunctions of $[T]$
but also closed under $[S]$, and the $N=\Nmax$ solutions
we have constructed are the only ones possible.
On a physical level, we can understand this result as follows.
Unlike the $(1,2)$ models in $D<10$,
our $(1,K)$ models are {\it in} their critical dimensions, and
therefore they lack ``internal'' right-moving degrees of freedom.
Thus, we expect that any dictionary relating a $K=2$ right-moving sector
to a $K>2$ right-moving sector should not take advantage of
these extra degrees of freedom in the $K=2$ sector in order to
introduce twists or (super)symmetry-breaking;  rather, we expect
these degrees of freedom to be ``frozen out'', handled jointly
as though they were one block.  This is precisely how they
appear in the dictionaries \dictionaryfour\ and \dictionaryeight,
and is the root of the previously encountered indistinguishability of
(or symmetry relating) the internal right-moving fermions.
Thus, it
is indeed sensible that our $(1,K)$ models are the analogues
of the $N=\Nmax$ $(1,2)$ models, for both are the unique models in which
no internal right-moving degrees of freedom are available for
(super-)symmetry breakings.  Note that this argument does not tell us the
degree of supersymmetry for the $(1,K)$ models themselves.
However, both an examination of the individual bosonic and fermionic degrees
of freedom and an overall counting of states
indicate that the $(1,K)$ strings have $N=1$ supersymmetry
[as distinguished from the $(K,K)$ strings, which have $N=2$
supersymmetry\refmark\letter].

Finally, we should point out that it is also possible to build
fractional superstring models which are {\it not}\/ spacetime-supersymmetric.
There are primarily two ways in which this can be done.
First, it is possible to construct string-function expressions
which are similar to $A_K$, $B_K$, and $C_K$ but which do {\it not}\/ vanish;
partition functions built with these expressions would then correspond
only to {\it non}-supersymmetric models, and analogous
dictionaries could be constructed (in the manner presented
in Sect.~IV) guaranteeing that such self-consistent $(1,K)$ models
actually exist.
Constructing such expressions is not difficult:  of all
the requirements listed after \expansion,
we need only eliminate the tachyon-free constraint $a_n=0$ for
all $n<0$.  Note that removal of this requirement need not
introduce spacetime tachyons
into the physical spectrum of a $(1,K)$ model,
for in general a partition function
$$   \calZ ~=~ \tautwo^k\,\sum_{m,n} \,a_{mn} \qbar^m\,q^n
\eqn\doubleexpansion
$$
will have physical tachyons only if a {\it diagonal}\/ element
$a_{nn}$ is non-zero for some $n<0$.  Indeed, it is easy to choose
holomorphic $K=1$ sectors for these models in such a way that
$a_{nn}=0$ for all $n<0$.

A second way to build non-supersymmetric partition functions
is to start with the supersymmetric
expressions $A_K$, $B_K$, and $C_K$ presented
in Sect.~III, but then separate them into their
spacetime bosonic and fermionic pieces in such a way that
$A_K=A_K^{(b)}-A_K^{(f)}$, \etc~~~
For the $K=2$
Jacobi identity $J=\gamma-\beta-\delta=0$, it is clear how
to do this:  the term $\beta$ represents the contributions
from Ramond (\ie, spacetime-fermionic) sectors in the theory,
and the remaining term $\gamma-\delta$ arises only
from a Neveu-Schwarz (\ie, spacetime-bosonic) sector.
For the $K>2$ ``para-Jacobi'' identities $A_K=B_K=C_K=0$, however,
the situation is more complicated.
A first approach might be to use the dictionaries developed in Sect.~IV
to relate our desired $K>2$ bosonic and fermionic pieces
to the known $K=2$ pieces:  we would simply split into such pieces
the spacetime Jacobi factor $J$
which appears in the $K=4$ and $K=8$ dictionaries,
and attempt to construct, for example, the $K=4$ string-function expressions
$A_4^{(b,f)}$ and $B_4^{(b,f)}$ which would be consistent with
the following dictionary:
$$  \eqalign{
       A_4^{(b)} ~&\equalstar~ 4k\,\deltaminushalf\,
              (\gamma-\delta)_{\rm st}\,(\gamma+\delta)_{\rm int} \cr
       A_4^{(f)} ~&\equalstar~ 4k\,\deltaminushalf\,
              (\beta)_{\rm st}\,(\gamma+\delta)_{\rm int} \cr
       B_4^{(b)} ~&\equalstar~ \phantom{4}k\,\deltaminushalf\,
              (\gamma-\delta)_{\rm st}\,(\beta+\gamma-\delta)_{\rm int} \cr
       B_4^{(f)} ~&\equalstar~ \phantom{4}k\,\deltaminushalf\,
              (\beta)_{\rm st}\,(\beta+\gamma-\delta)_{\rm int} \cr
     }
\eqn\fourpieces
$$
where we have explicitly indicated with subscripts the separate
$K=1$ spacetime and internal factors.
However, it is easy to see that this method is not appropriate,
for we do not expect the cancellations occurring in the full
expressions $A_K$ and $B_K$ to mirror the relatively simple
cancellation occurring in the $K=2$ case.  Indeed, it is easy
to show that no string functions for the left sides of \fourpieces\
can be found which transform under $[S]$ and $[T]$ as do
the right sides.  Instead, one can determine the separate bosonic
and fermionic contributions to the expressions $A_K$, $B_K$ and $C_K$
by demanding that these individual contributions each have
$q$-expansions $\sum_n a_n q^n$ in which all
coefficients $a_n$ are non-negative, and have relatively
simple closure relations under $[S]$.  This approach, in fact,
proves successful, and yields results consistent with our interpretation
[discussed after \ABCKequaleight]
that the terms $(c^0_0)^{D-3}(c^2_0)$ are bosonic and
the terms $(c^{K/2}_{K/2})^{D-2}$ are fermionic.\refmark{\ADT}\
Thus, the construction
of {\it non}-supersymmetric $(1,K)$ models is in principle
no more difficult than that of the supersymmetric models we have already
considered, and we expect our dictionary techniques to
generalize to these cases as well [though of course not yielding
dictionaries similar to \fourpieces].

To summarize the results of this and the previous section, then,
we have succeeded in developing a method by which
the partition functions of valid supersymmetric $(1,K)$ models
can be generated and their gauge groups identified.
This was achieved by establishing a correspondence between
$(1,K)$ models and $(1,2)$ models in $D=2+16/K$ dimensions,
resulting in two dictionaries [Eqs.~\dictionaryfour\ and
\dictionaryeight] enabling one to ``translate'' or ``substitute'' between
the understood $K=2$ sector and the less-understood spacetime-supersymmetric
$K>2$ sector.
We found that only $(1,2)$ models with $N=\Nmax$ SUSY
were translatable, and we were thereby able to estimate the
sizes of the spaces of $(1,K)$ models.

\endpage

\chapter{FURTHER DISCUSSION AND REMARKS}

In this concluding section we discuss
two different extensions of our results:
these are the questions of obtaining chiral fermions and
achieving Lorentz invariance.
We begin by investigating
how chiral fermions -- or more generally, chiral supermultiplets --
might arise in the fractional superstring models we have
been investigating.  We discuss several different methods
for obtaining such multiplets, one of which involves formulating
or interpreting these models in dimensions less than their critical
dimensions.  While this might seem to spoil the original
attraction of the fractional superstring approach, we find
instead that such a compactification is not at all arbitrary
(as it is for the traditional superstrings),
but rather is required in order to achieve a self-consistent
Lorentz-invariant interpretation.
Indeed, we find that requiring Lorentz invariance seems
to specify a ``natural'' dimension in which the theory
must be formulated, thereby (in a unique manner) simultaneously offering a
possible solution to the chiral fermion problem.
We emphasize that such a ``forced'' compactification appears
to be a feature wholly new to fractional superstrings.
Furthermore, we find that the ``natural'' dimension for the $K=4$ string
appears to be $D=4$, rendering the $K=4$ fractional superstring
the most likely candidate for
achieving chiral particle representations in four-dimensional
spacetime while maintaining Lorentz invariance.

We begin by investigating
how we might obtain chiral massless spacetime
supermultiplets (\ie, supermultiplets which transform
in a complex representation of the gauge group)
in our fractional superstrings.
Recall from Sect.~III that
the massless particles appear only in the $A_K$ sectors
of the partition functions we have examined;
indeed, for $K\geq 2$,
the $(c^0_0)^{D_c-3} (c^2_0)$ term within $A_K$ contains the
contributions of the right-moving components of the massless
vector particles $A^\mu$ while the $(c^{K/2}_{K/2})^{D_c-2}$
term contains those of the massless fermion $\psi^\alpha$
(where $\alpha$ is a spacetime Lorentz spinor index).
For a heterotic $(1,K)$ string theory in its critical
dimension, the massless spacetime supermultiplet structure is
in general
$$      (\psi^\alpha,A^\mu)_R ~\otimes~
       \biggl\lbrace
    X^\nu ~\oplus ~\phi^{(h_1)} ~\oplus ~\phi^{(h_2)} ~\oplus~...
     \biggr\rbrace_L~;
\eqn\oneKmultiplet
$$
here $(\psi^\alpha,A^\mu)_R$ is the fermion/vector supermultiplet
discussed above, and for the
left-moving excitations we have indicated the various possible massless
states:  $X^\nu$ denotes the Lorentz-vector (gauge-singlet) state
achieved by exciting the worldsheet
left-moving component of the spacetime coordinate boson,
and the $\phi^{(h_i)}$ denote the various remaining Lorentz-scalar
states combined into representations
(with highest weights $h_i$) of the relevant left-moving gauge group.
Within \oneKmultiplet, the combination
$$    (\psi^\alpha,A^\mu)_R ~\otimes ~(X^\nu)_L
\eqn\supergravmultiplet
$$
forms the usual $N=1$ supergravity mutliplet
containing the
spin-2 graviton $g^{\mu\nu}$, spin-1 antisymmetric tensor field
$B^{\mu\nu}$, spin-0 dilaton $\phi$, spin-3/2 gravitino
$\lambda^{\nu\alpha}$, and spin-1/2 fermion $\lambda^\alpha$.
All of these states are of course gauge-singlets.
The only other combination within \oneKmultiplet\
allowed by level-matching constraints is
$$  (\psi^\alpha,A^\mu)_R ~\otimes ~(\phi^{(h_{\rm adj})})_L
\eqn\adjointonly
$$
where these left-moving states fill out the adjoint representation
of the gauge group.  This yields, of course, a spacetime supermultiplet
consisting of spin-1 gauge bosons $A^\mu_{\rm adj}$ and
their spin-1/2 fermionic superpartners $\psi^\alpha_{\rm adj}$.
The crucial point, however, is that this supermultiplet structure
therefore does not permit spacetime fermions which
transform in any other (\eg, complex) representations
of the gauge group.

In the usual superstring phenomenology, a chiral supermultiplet
can be achieved by introducing an
additional right-moving supermultiplet of the form $(\phi_i,\psi^\alpha)_R$
where the $\phi_i$, $i=1,...,D-2$, are a collection of Lorentz scalar fields.
We then obtain, as in \oneKmultiplet, the additional states
$$      (\phi_i,\psi^\alpha)_R ~\otimes~
       \biggl\lbrace
   X^\nu ~\oplus ~\phi^{(h_1)} ~\oplus~\phi^{(h_2)} ~\oplus~...
     \biggr\rbrace_L~.
\eqn\oneKmultipletscalar
$$
The combination
$$   (\phi_i,\psi^\alpha)_R ~\otimes~ (X^\nu)_L
\eqn\Ngrow
$$
yields additional spin-1 vector bosons $\tilde{A}_i^\nu$, spin-1/2 fermions
$\psi^\alpha$, and spin-3/2 gravitinos
$\lambda^{\nu\alpha}$ (thereby producing an $N>1$ SUSY unless
these extra gravitinos are GSO-projected out of the spectrum).
However, since \oneKmultipletscalar\
is built with the right-moving supermultiplet
$(\phi_i,\psi^\alpha)_R$, the level-matching constraints now
also allow the additional combinations
$$  (\phi_i,\psi^\alpha)_R ~\otimes~ (\phi^{(h_i)})_L
\eqn\chiralsupermultiplet
$$
where the gauge-group representations are {\it not}\/
restricted to the adjoint.
Thus, these combinations \chiralsupermultiplet\
can yield chiral spacetime supermultiplets when $h_i$ are the
highest weights of complex representations;
in particular, \chiralsupermultiplet\ can indeed contain chiral massless
spacetime fermions transforming in the gauge-group fundamental representation.

How might these additional supermultiplets $(\phi_i,\psi^\alpha)_R$ arise
in our fractional superstring theories?
At the partition function level, we can simply introduce new terms
of the form
$(c^0_0)^{D_c-3} (c^2_0)$ to be interpreted as containing
the contributions of massless scalar fields.  Such $(\phi_i,\psi^\alpha)$
multiplets
could then be accommodated simply by multiplying the overall $K=4$ or $K=8$
partition functions by an appropriate integer (so as to maintain
modular invariance).  At the
level of the actual particle spectrum, however, it is not clear
how such extra fields might arise.
Extra fermions are not difficult to find, since
we start with $2^D$ degrees of freedom while the Dirac $\gamma$-matrix
algebra, along with the Majorana/Weyl condition, reduces this to $2^{D/2-1}$
degrees of freedom.  This clearly leaves many other degrees of freedom
remaining for chiral supermultiplets.  Extra scalar fields,
on the other hand, are more difficult to
obtain;  one possible scenario is as follows.
For $K>2$
there also exist extended parafermion
theories;\REF\ed{P.C.~Argyres, E.~Lyman, and S.-H.H.~Tye,
 {\it Low-lying States of the Six-Dimensional Fractional Superstring},
Cornell preprint CLNS 91/1121 (to appear).}\refend\
these have the same central charge as our usual parafermion theories,
but contain more than one $\phi^1_0$ field.
At first sight, these extra $\phi^1_0$ fields will have
spacetime Lorentz indices $\mu$ associated with them.
However, recall
that in the parafermion theories there are additional parafermion fields
whose characters have not appeared in the partition functions:  these
are the half-integer spin fields.
It may therefore be possible to interpret these extra $\phi^1_0$
fields as Lorentz scalar {\it composites} of the half-integer spin fields --
\eg, the composite $(\phi^{1/2}_{1/2})^\mu (\phi^{1/2}_{-1/2})_\mu$
may contain the additional $\phi^1_0$ field in the extended
parafermion theory.  Forming these $\phi^1_0$ fields as composites
may allow them to be interpreted as Lorentz scalars, much as the scalar
composite $:\epsilon^\mu
\epsilon_\mu:$ appears as one of the descendants of $\epsilon^\mu$
[as in \epsepsOPE].
These scalars would then supply the scalar fields needed for chiral
supermultiplets.
Of course, a detailed analysis is necessary to see if these scenarios
are possible.

A more widely-known and established method of generating supermultiplets
$(\phi_i,\psi^\alpha)$ in string theory is through spacetime compactification.
Since the $K=8$ string
is already in $D_c=4$, let us focus on the $(1,4)$ heterotic string with
$D_c=6$.
For this string, we can choose to compactify two space dimensions:
$$  (\psi^\alpha,A^\mu)~\longrightarrow~
    (\psi^\alpha,A^\mu) ~\oplus~ (\phi_i,\psi^\alpha)~;
\eqn\compactify
$$
here the $A^\mu$, $\mu=0,1,...,5$ in the $D=6$ theory breaks
down to $A^\mu$, $\mu=0,1,2,3$ in $D=4$, with the remaining
fields reidentified as internal scalars:  $A_4=\phi_1$ and $A_5=\phi_2$.
Similarly, the four-component Weyl fermion in $D=6$ splits
into two two-component fermions in $D=4$.
Therefore, the degree-of-freedom count corresponding
to \compactify\ is
$$    (4 +4) ~\longrightarrow~ (2+2) ~+~ (2+2)~.
\eqn\dofcount
$$
In the usual $(1,2)$ string compactified to four spacetime dimensions,
it is precisely this mechanism which
breaks the ten-dimensional Lorentz symmetry to
four-dimensional Lorentz symmetry.
The separate multiplets
$(\psi^\alpha,A^\mu)_R$ and $(\phi_i,\psi^\alpha)_R$ can then
couple to different left-moving sectors, allowing the possibility of
chiral fermions in the fundamental representation of the relevant gauge group.

There is, however, a crucial difference between the compactification
of the usual $(1,2)$ heterotic string and the compactification of
the $(1,4)$ string to four spacetime dimensions.  In the usual
superstring, the compactification is {\it ad hoc}:  one arbitrarily
chooses the resulting dimension in which one wishes to formulate the $(1,2)$
theory, and there are neither dynamical nor symmetry reasons
why four dimensions is chosen.  For the $(1,4)$ string, however,
we can argue that even though the critical dimension is $D_c=6$,
the partition function itself indicates that there does not exist
a six-dimensional Minkowskian spacetime interpretation consistent with
Lorentz invariance, locality, and quantum mechanics.
Indeed, we will see that the largest number of spacetime dimensions
in which a consistent interpretation exists is $D=4$.  We shall call
this the {\it natural dimension}.  Hence, the
$(1,4)$ string appears unique in that its compactification
to (or interpretation in)
four spacetime dimensions is essentially induced by the intrinsic
structure of the theory itself.

Let us see how this comes about.
Recall that $c^\ell_n=c^{2j}_{2m}$ is the character for
$[\phi^j_m]$.  Also recall that the parafermion fusion rules \fusionrules\
indicate that the $m$-quantum number is additive (modulo $K/2$)
while the $j$-quantum number is not.
As we have already observed, the $(c^0_0)^3(c^2_0)$ term
in $A_4$ represents the contributions from spacetime bosons,
and the $(c^2_2)^4$ term in $A_4$ represents the contributions
from spacetime fermions.  Hence, as is argued
in Ref.~[\ADT], terms with
$D-2$ factors of $c^{\ell}_{K/2}$ (with arbitrary values of $\ell$)
correspond to spacetime fermions, while terms of the forms
$(c^{\ell}_0)^{D-2}$ (with arbitrary values of $\ell$)
correspond to spacetime bosons.
Such identifications arise essentially from recognizing the
worldsheet vacuum state corresponding to $(\phi^0_0)^{D-2}$
as yielding spacetime bosons
[it is the analogue of the $K=2$ Neveu-Schwarz vacuum $(\phi^0_0)^8$],
and recognizing the worldsheet vacuum state corresponding
to $(\phi^{K/4}_{\pm K/4})^{D-2}$
as yielding spacetime fermions [it is the analogue of the
$K=2$ Ramond vacuum $(\phi^{1/2}_{\pm 1/2})^8$].
Note that
this identification is in fact consistent at all mass levels,
since excitations upon these respective vacua occur only through the
$\epsilon=\phi^1_0$ field, and the parafermion fusion rules show
that applications
of $\phi^1_0$ cannot change the $m$-quantum numbers of these
respective vacua.  Furthermore,
since the $m$-quantum number is additive modulo $K/2$,
this identification indeed reproduces the desired Lorentz
fusion rules $B\otimes B=B$, $B\otimes F=F$, and $F\otimes F=B$,
where $B$ and $F$ represent spacetime bosonic and fermionic fields.
Thus, with this interpretation in $D=D_c$ spacetime dimensions, we
see that it is straightforward
to determine the spacetime statistics of the particles
contributing to each term in $A_4$ and $A_8$.

However, we now note that in $B_4$ there
exists a term of the form $(c^2_2)^2(c^2_0)^2$ which has only
massive states [\ie, in a $q$-expansion of this term we
find $q^h(1+...)$ where $h>0$].
Although this term
may have more than one interpretation, the elementary
particles giving rise to this term cannot be interpreted
as either bosons or fermions in six spacetime dimensions;
indeed,
this term cannot arise through $\epsilon$-excitations from either
the six-dimensional bosonic or fermionic vacuum state.
It is, therefore, natural to interpret this term as the contribution
from either:
\item{1.}  fermions in four spacetime dimensions [thereby
   explaining the $(c^2_2)^2$ factor] with additional internal
   quantum numbers [yielding the $(c^2_0)^2$], or
\item{2.}  bosons in four spacetime dimensions [thereby
   explaining the $(c^2_0)^2$ factor] with additional internal
   quantum numbers [yielding the $(c^2_2)^2$].

How might such internal factors arise?
Let us recall the case of the usual superstring theory,
where in $D_c=10$ spacetime dimensions the Ramond sector
(corresponding to spacetime fermions) always contributes
to the partition
function the factor $-(c^1_1)^8=-\thetatwo^4/(16\deltahalf)$.
Indeed, this is the term appearing in $A_2$ in Eq.~\AKequaltwo;
the presence of the eighth-power of $c^1_1$ is a reflection
of the underlying rotational invariance of the eight transverse
directions.  However, compactifying a model to a dimension
$D<D_c$ reduces the power required;  indeed, we have seen
in Sect.~III that for models compactified to $D<D_c$
flat spacetime dimensions, we obtain terms in the partition
function with string-function factors of the form $c^{D-2} c^{D_c-D}$
where the first factor
comes from spacetime-associated degrees of freedom (\ie, the worldsheet
parafermions carrying a spacetime Lorentz index) and where the second
factor comes from additional internal degrees of freedom (\ie,
worldsheet parafermions without such indices).
In the present case, upon compactification we would expect terms
of the form $(c^\ell_{K/2})^{D-2} (c^{\ell'}_{0})^{D_c-D}$
where the second factor represents contributions from spacetime
bosons (and is the $K>2$ analogue of the contributions from
the usual Neveu-Schwarz sectors of the theory).
Remarkably, this
is precisely what occurs in $B_4$.  Apparently, six-dimensional
Lorentz invariance for the $K=4$ string appears to be explicitly
broken.

Note that -- at tree-level --
the particles in the $A_4$ sector couple only
to other fields in the $A_4$ sector;
in particular, the tree-level scatterings of massless particles do
not involve the $B_4$ sector.
At the string {\it loop} level, however,
fields in the $B_4$ sector
couple to the massless particles in the $A_4$ sector, so we see
that it is the string's own quantum effects which render the
theory in its critical dimension inconsistent with Lorentz invariance.
Thus, in this sense, it is quantum mechanics
and the requirement of Lorentz invariance which together necessitate
or {\it induce} the compactification from the critical dimension (six)
to the natural dimension (four).
This mechanism is quite remarkable,
and has no analogue in the traditional $K=1,2$ string theories
(in which the critical dimensions {\it are} the natural dimensions).
In particular, only for the $K>2$ theories do such independent $B$-sectors
appear;  for $K=2$ there is only one possible supersymmetric
$A$ sector, corresponding to the one supersymmetry identity
(Jacobi identity) for $K=2$.  It is the presence of a unique $B_K$ sector
for $K>2$, itself necessitated by modular invariance and closure
under $[S]$, which induces this compactification.

The same argument applies to the $K=8$ string (for which the critical
dimension is four but the natural dimension is three),
but with one additional feature.
As before, the $A_8$ expression [Eq.~\ABCKequaleight]
in the $K=8$ string
is consistent with a $D=D_c=4$ spacetime
boson/fermion interpretation.  However, we can quickly see
that the $B_8$ expression is {\it not} consistent with this interpretation;
rather, the form of this expression
dictates a {\it three}-dimensional spacetime interpretation.
Now, the {\it third}\/ expression $C_8$ in \ABCKequaleight\ contains string
functions only of the type $c^\ell_{K/4}=c^\ell_2$.
(Recall that $c^\ell_6=c^{K-\ell}_2$ for $K=8$.)
Not only is $C_8$ therefore consistent with three-dimensional
Lorentz invariance (and in fact Lorentz invariance in any number
of spacetime dimensions), but our above interpretation indicates
that $C_8$  must correspond
to fields with fractional spacetime Lorentz spins $1/4$ and $3/4$.
However, as is well-known,
fractional-spin fields (\ie, anyons) are consistent with quantum mechanics and
Lorentz invariance in three spacetime dimensions, thereby providing
an independent way in which self-consistency selects this dimensionality.
Thus, once again our $K>2$ strings seem
to select naturally their proper spacetime dimensions:  these
are the dimensions in which consistent Lorentz-invariant interpretations
can be achieved.

Note that
if such speculations are indeed correct, then the $(1,4)$
and $(1,8)$ models we have constructed in this paper
have continuous (and non-chiral) spectra, for when we interpret them in
four and three spacetime dimensions respectively these string
theories contain worldsheet bosons which remain uncompactified.
However, we expect the ``dictionary'' approach we have developed in this
paper to be easily generalizable to the compactified (and chiral)
case as well.

In summary, then, although there may well exist other possible
interpretations of these partition functions, we find the above
speculations both tantalizing and intriguing.
It is indeed fortuitous that the solution to what might have
seemed a problem (\ie, finding an interpretation consistent with Lorentz
invariance) also simultaneously provides a possible solution to the more
phenomenological problem of obtaining models with chiral fermions.
It is also strongly compelling that the quantum structure
of the theory itself seems to dictate this solution.
These are issues clearly worth investigating.

\endpage

\ACK
We are pleased to thank S.-W.~Chung, J.~Grochocinski,
O.~Hern\'andez, C.-S.~Lam, E.~Lyman, R.~Myers, and especially
P.~Argyres for many useful discussions.  K.R.D.~also thanks the
High-Energy Theory Group at Cornell University for its hospitality while
portions of this work were carried out.
This work was supported in part by the U.S.~National
Science Foundation,
the National Science and Engineering Research Council of Canada, and
les Fonds FCAR du Qu\'ebec.

\endpage

\Appendix{A}

In this appendix we introduce the string functions (or parafermion
characters) $c^\ell_n$, collecting together their definitions,
modular transformation properties, special cases, and identities.
The results quoted here are
due mostly to Ka\v{c} and Peterson.\refmark{\kacpeterson}\

The string functions are essentially the characters $Z^{2j}_{2m}$
of the $\bZ_K$ parafermion
fields $\phi^j_m$:
$$     Z^\ell_n ~=~ \eta\,c^\ell_n
\eqn\appear
$$
where $\ell\equiv 2j$, $n\equiv 2m$, and where $\eta$ is the Dedekind
$\eta$-function:
$$    \eta(\tau) ~\equiv~ q^{1/24} \,\prod_{n=1}^\infty (1-q^n) ~=~
         \sum_{n= -\infty}^\infty \,(-1)^n\,q^{3(n-1/6)^2/2} ~
\eqn\Dededef
$$
with $q\equiv e^{2\pi i\tau}$.  Since the $\bZ_K$ parafermion theory
can be represented as the coset theory
$SU(2)_K/U(1)$, the string functions can
therefore be obtained by expanding the full $SU(2)_K$ characters
$\chi_\ell(\tau,z)$
in a basis of $U(1)$ characters $\Theta_{n,K}(\tau,z)/\eta$:
$$   \chi_\ell(\tau,z) ~=~ \sum_{n= -\ell}^{2K-\ell-1} \,
       Z^\ell_n (\tau) ~{{\Theta_{n,K}(\tau,z)}\over \eta}~
      ~=~ \sum_{n= -\ell}^{2K-\ell-1} \,
       c^\ell_n (\tau) \,\Theta_{n,K}(\tau,z)~.
\eqn\basisexpansion
$$
Here $\Theta_{n,K}(\tau,z)/\eta$ are the characters of a $U(1)$ boson
compactified on a radius $\sqrt{K}$.
Since the $SU(2)_K$ characters and the $\Theta_{n,K}$ functions are
well-known, explicit expressions
for the string functions can be extracted.
Eq.~\basisexpansion, then, can be taken as a definition of the
string functions $c^\ell_n$.

Expressions for
the string functions $c^\ell_n$ were originally obtained
by Ka\v{c} and Peterson;\refmark{\kacpeterson}\
for our purposes, however, a useful and equivalent expression
is that due to Distler and
Qiu:\REF\Jacques{J.~Distler and Z.~Qiu, {\it Nucl.~Phys.~}{\bf B336},
533 (1990).}\refend\
$$  \eqalign{
  c^\ell_n(\tau) ~&\equiv~ q^{h^\ell_n +
        [4(K+2)]^{-1}} \,\eta^{-3}\,
      \sum_{r,s=0}^\infty
   \, (-1)^{r+s}\,q^{r(r+1)/2\,+\,
      s(s+1)/2\,+\,rs(K+1)}\,\times\cr
 &~~\times \biggl\lbrace
     q^{r(j+m)\,+\, s(j-m)}~
     -~ q^{K+1-2j\,+\,r(K+1-j-m)\,+\,s(K+1-j+m)}\biggr\rbrace\cr}
\eqn\stringdefinition
$$
where $\ell-n\in 2{\bf Z}$
and where the highest weights $h^\ell_n$ are
$$  h^\ell_n ~\equiv~ {{\ell(\ell+2)}\over{4(K+2)}} ~-~{{n^2}\over{4K}} ~~~~~~
    {\rm for~} |n|\leq \ell ~.
\eqn\hdef
$$
The string functions have the symmetries
$$    c^\ell_n ~=~ c^{K-\ell}_{K-n} ~=~ c^\ell_{-n} ~=~ c^\ell_{n+2K}~,
\eqn\stringshifts
$$
as a consequence of which for any $K$ we are free to choose a ``basis''
of string functions $c^\ell_n$ where $0\leq \ell\leq K$ and
$0\leq n \leq n_{\rm max}$ [where $n_{\rm max}$ equals $\ell$
if $\ell\leq K/2$, and $\ell-2$ otherwise].

The string functions are closed under the modular transformations
$S$ and $T$.   In fact, under $T:\tau\to\tau+1$ they transform
as eigenfunctions:
$$ c^\ell_n(\tau+1)~=~ \exp\lbrace 2\pi i s^\ell_n \rbrace~c^\ell_n(\tau)
\eqn\Ttrans
$$
where the phase $s^\ell_n$ is given by
$$   s^\ell_n ~\equiv~ h^\ell_n ~-~ {1\over{24}}\,c_0 ~=~
      h^\ell_n ~-~ {{K}\over{8(K+2)}} ~.
\eqn\Tphase
$$
Here $h^\ell_n$ is defined in \hdef, and $c_0$ is the
central charge of the full $SU(2)_K$ theory [given in \czero].
Under $S:\tau\to -1/\tau$, the level-$K$ string functions
mix among themselves:\refmark{\kacpeterson}\
$$ c^\ell_n(-1/\tau) ~=~ {1\over{\sqrt{-i\tau}}}\,{1\over{\sqrt{K(K+2)}}}\,
      \sum_{\ell'=0}^K ~
     \sum^K_{{n'= -K+1}\atop{\ell^\prime-n^\prime\in 2{\bf Z}}}
      \,b(\ell,n,\ell',n') \,c^{\ell'}_{n'}
\eqn\Strans
$$
where the first square root indicates the branch with non-negative
real part and
where the mixing coefficients $b(\ell,n,\ell',n')$ are
$$  b(\ell,n,\ell',n') ~\equiv~ \exp\left\lbrace
    {{i\pi nn'}\over{K}} \right\rbrace \,\sin\left\lbrace
     {{\pi (\ell+1)(\ell'+1)}\over{K+2}} \right\rbrace ~.
\eqn\bdef
$$
 From \Ttrans\ and \Strans\ it follows that the string functions
have modular weight $k= -\half$.

Note that in
the special $K=1$ and $K=2$ cases, the string functions can be expressed
in terms of the Dedekind $\eta$-function \Dededef\ and
the more familiar Jacobi $\vartheta$-functions;  these
relations are given in \cforKequalone\ and \cforKequaltwo.
In particular, these $\vartheta$-functions are
$$  \eqalign{
      \thetatwo(\tau) ~&\equiv~ 2\,q^{1/8}\,\prod_{n=1}^\infty\,
                 (1+q^n)^2\,(1-q^n) ~\cr
      \thetathree(\tau) ~&\equiv~ \prod_{n=1}^\infty\,
                 (1+q^{n-1/2})^2\,(1-q^n) ~\cr
      \thetafour(\tau) ~&\equiv~ \prod_{n=1}^\infty\,
                 (1-q^{n-1/2})^2\,(1-q^n) ~,\cr }
\eqn\thetadefs
$$
and under the $[S]$ and $[T]$ transformations
[where the bracket indicates the stroke operator, defined in \strokedef],
we find
$$  \eqalign{
    \thetatwo[S]~=~ \exp(7\pi i/4)\,\thetafour~,&~~~~~
    \thetatwo[T]~=~ \exp(\pi i/4)\,\thetatwo~,\cr
    \thetathree[S]~=~ \exp(7\pi i/4)\,\thetathree~,&~~~~~
    \thetathree[T]~=~ \thetafour~,\cr
    \thetafour[S]~=~ \exp(7\pi i/4)\,\thetatwo~,&~~~~~
    \thetafour[T]~=~ \thetathree~,\cr
    \eta[S]~=~ \exp(7\pi i/4)\,\eta\phantom{_2}~,&~~~~~
    \eta[T]~=~ \exp(\pi i/12)\,\eta~.\cr}
\eqn\thetatransforms
$$
Note that \thetatransforms\ is indeed consistent with the
string-function transformations \Ttrans\
and \Strans\ for $K=2$;  in particular, under $[S]$ we find that
the combination $c^0_0+c^2_0$ transforms as an eigenfunction
while $c^0_0-c^2_0$ and $c^1_1$ transform into each other.

The $S$-transformation formula \Strans\ assumes an analogously
simple form for the $K=4$ case.
In this case there are seven distinct string functions,
and under $[S]$ they can be block-diagonalized as follows:
$$   \pmatrix{ c^0_0+c^4_0 \cr c^2_0 \cr c^4_2 \cr
     c^2_2 \cr} \,[S] ~=~ e^{i\pi/4}\,{\bf R^+} \,
   \pmatrix{
     c^0_0+c^4_0 \cr c^2_0 \cr c^4_2 \cr c^2_2 \cr}
\eqn\mixingsone
$$
and
$$   \pmatrix{
       c^0_0-c^4_0 \cr c^1_1 \cr c^3_1 \cr }
      \,[S]~=~ e^{i\pi/4}\,{\bf R^-}\,
      \pmatrix{
       c^0_0-c^4_0 \cr c^1_1 \cr c^3_1 \cr }
\eqn\mixingstwo
$$
where
$$   {\bf R^+} ~\equiv~
      {1\over{\sqrt{24}}} \,\pmatrix{
      2 & 4 & 4 & 4 \cr 2 & -2 & 4 & -2 \cr 1 & 2 & -2 & -2 \cr
      2 & -2 & -4 & 2 \cr }~,~~~~~~
     {\bf R^-} ~\equiv~
      {1\over{\sqrt{24}}} \,\pmatrix{ 0 & 4\sqrt{3} & 4\sqrt{3} \cr
      \sqrt{3} & \sqrt{6} & -\sqrt{6} \cr
        \sqrt{3} & -\sqrt{6} & \sqrt{6} \cr } ~.
\eqn\SmatricesRS
$$
Note that $({\bf R^+})^2=({\bf R^-})^2={\bf 1}$, as required.

This mixing pattern seen for the $K=2$ and $K=4$ cases
exists for other higher even values of $K$ as well.  For any such $K$,
we can define the linear combinations
$d^{\ell\pm}_n\equiv c^\ell_n\pm c^{K-\ell}_n$ when $(\ell,n)$ are even;
note that these $d$-functions will also be eigenfunctions of $[T]$
if $K\in 4{\bf Z}$ because in these cases
$s^\ell_n=s^{K-\ell}_n$ (mod 1).  It then follows from \Strans\ that
the $d^+$-functions mix exclusively among themselves under $S$
[as in \mixingsone], and that the $d^-$-functions mix
exclusively with themselves and with the odd $(\ell,n)$
string functions [as in \mixingstwo].
It is for this reason that (for $K>2$)
it is possible to construct modular-invariant
expressions involving the string functions in such a manner that
the odd string functions do not appear.  Indeed, one can check
that in all the string-function expressions which have appeared
for the fractional superstring (such as $A_K$, $B_K$, and $C_8$),
only the $d^+$ combinations have played a role.

\endpage

\Appendix{B}

In this appendix we demonstrate, using the free-fermionic
spin-structure construction of Ref.~[\KLT], that
any $(1,2)$ model in $D<10$ spacetime dimensions with a partition
function of the form
$$   \calZ ~=~ \tautwo^{-2}\, \deltaminusone\,\deltabarminushalf\,\Jbar\,
       \biggl\lbrace\bbar^{(10-D)/4} X + \cbar^{(10-D)/4} Y +
    \dbar^{(10-D)/4} Z\biggr\rbrace ~
\eqn\translatableagain
$$
has an $N=\Nmax$ spacetime supersymmetry, where $\Nmax$ is given in \nmax.
We assume the reader to be familiar with the model-construction
procedure described in Ref.~[\KLT], and we use the same notation
here.

It is clear that a model with partition function \translatableagain\
is space\-time-su\-per\-sym\-met\-ric;  therefore, as discussed in Ref.~[\KLT],
its set of spin-structure generating vectors $\W{i}$ must include
the two vectors
$$  \eqalign{
     \W{0}~&\equiv~ [(\half)^{12+2k} ~|~ (\half)^{24+2k}]~\cr
   {\rm and~~~~}
      \W{1}~&\equiv~ [(0)^{-k} (0\half\half)^{4+k} ~|~ (\half)^{24+2k}]~,\cr}
\eqn\spinstructure
$$
where $k$, the modular weight, equals $1-D/2$.
Here the left sides of the ${\bf W}$-vectors indicate the boundary
conditions of the {\it right}-moving complex worldsheet fermions,
and we see that the four complex fermions with Ramond boundary conditions
in $\W{1}$ (\ie, the four whose components in $\W{1}$ are zero)
are the ones which together contribute the spacetime Jacobi
factor $\Jbar$ in \translatableagain.
The remaining $2(4+k)=10-D$ right-moving complex worldsheet fermions
(\ie, those whose components in $\W{1}$ equal $\half$) are therefore
the ones producing the remaining factors of $\beta^{(10-D)/4}$,
$\gamma^{(10-D)/4}$, and $\delta^{(10-D)/4}$ in \translatableagain.

In order for our model to have a partition function of the
form \translatableagain,
these remaining $(10-D)$ worldsheet fermions
must have the {\it same} toroidal boundary conditions in
all sectors of the model which contribute to the partition
function.  In particular, this must be true {\it separately}
for spacetime bosonic and fermionic
sectors (\ie, those which have $\alpha s=\half,0$
respectively in the notation of Ref.~[\KLT]).
This can occur only if every other spin-structure generating
vector $\W{i}$ has equal components for these $10-D$ fermions
(thereby giving rise to the maximal right-moving ``gauge'' symmetry discussed
in Sect.~V).
Thus, any other spin-structure vectors $\W{i}$ in the generating
set must have a right-moving component of the form
$$    \W{i}^R~=~ [(0)^{-k} \,(X)^{4+k} ]
\eqn\twochoices
$$
where each factor $X$ is independently either $(000)$ or $(0\half\half)$.
In determining \twochoices\ we
have had to satisfy the ``triplet'' constraint;  we have
also made use of our freedom to choose (without loss of
generality) vectors which have their first components vanishing.

Let us now consider the constraints which must be satisfied by
the gravitino states.
These states all arise in the $\W{1}$ sector, and have
the charge vectors
$$    {\bf Q}_{\rm grav}~\equiv~{\bf N}_{\W{1}}+\W{1}-\W{0}~=~
         [(\pm \half)^{-k} (\pm \half 00)^{4+k} ~|~ (0)^{24+k}]
\eqn\chargevectors
$$
where in principle all $\pm\half$ combinations are allowed.
The constraint due to the $\W{0}$ vector, however, immediately
projects out half of these states, allowing only those with charge
vectors in which the product of non-zero components is positive.
This by itself leaves us with eight distinct allowed states,
and these states are precisely those which combine to
form the original $\Nmax$ gravitinos.  Thus, our model
will contain {\it fewer} gravitinos
if and only if the other ${\bf W}$-vectors in the generating
set produce constraint equations removing some of these gravitinos
from the physical spectrum.
The $\W{1}$-vector provides no further constraint, however, and
since any additional ${\bf W}$-vectors must have right-moving
components of the forms \twochoices, they each give rise to
constraint equations of the form
$$   0 ~=~f(k_{ij})
\eqn\otherconstraints
$$
where $f(k_{ij})$ is a function of the projection constants $k_{ij}$
(defined in Ref.~[\KLT]).
Note, in particular, that \otherconstraints\ does not involve
the charge vectors \chargevectors.
Thus, we see that either {\it all}\/ or {\it none} of the $\Nmax$
gravitinos remain in the physical spectrum:  all gravitinos
remain if the projection constants $k_{ij}$ are chosen
so that in each case $f(k_{ij})=0$, and none remain otherwise.
However, we know that our model must have at least an
$N=1$ spacetime supersymmetry.
Therefore, in order to have a partition function of
the form \translatableagain,
our model must have $N=\Nmax$ spacetime supersymmetry.

\endpage
\refout
\bye